\documentclass[reprint, amsmath, amssymb, aps, prx, superscriptaddress]{revtex4-2}
\usepackage{graphicx} 
\usepackage{braket} 
\usepackage{bm}
\usepackage{booktabs}
\usepackage{array}
\newcolumntype{C}[1]{>{\centering\arraybackslash}p{#1}}
\newcolumntype{L}[1]{>{\raggedright\arraybackslash}p{#1}}
\usepackage{tikz}
\usepackage{pgfplots}
\usepackage{xcolor}
\usepackage{colortbl}
\usepackage{bbm}
\usepackage{amsfonts}
\usepackage{mathtools}
\usepackage{LVcolours} 
\usepackage{appendix}
\usepackage{algpseudocode}
\usepackage[hidelinks, colorlinks=true]{hyperref}

\begin{document}

\title{Flag at origin: a modular fault-tolerant preparation for CSS codes}

\author{Diego Forlivesi}
\affiliation{Quantinuum, Terrington House, Cambridge, CB2 1NL, United Kingdom}
\author{David Amaro}
\affiliation{Quantinuum, Partnership House, Carlisle Place, London SW1P 1BX, United Kingdom}

\date{\today}

\begin{abstract}
Fault-tolerant (FT) preparation of diverse logical stabilizer states in quantum error-correcting (QEC) codes is essential for FT computation.
Existing constructions of these FT circuits are often constrained by classical computational resources or result in unnecessarily large quantum circuits.
This work introduces a modular construction for FT preparation circuits in CSS codes of arbitrary distance, yielding significantly more resource-efficient circuits than previous approaches—especially for the largest codes studied.
The key insight is that in bipartite CX circuits used to prepare CSS states, $X$ errors propagate from the partition of \textit{control qubits} to the partition of \textit{target qubits}, while $Z$ errors propagate in the opposite direction.
By appending $X$-detecting flag gadgets to the control partition and $Z$-detecting flag gadgets to the target partition, the circuit becomes FT.
To manage the associated overhead, we propose an algorithm that discovers optimal (or near-optimal) flag gadgets at any distance.
These gadgets are reusable across different QEC codes and FT subroutines, such as flag-based QEC.
We estimate the logical state preparation error using subset-sampling Monte Carlo simulations at the circuit level, combined with approximate maximum-likelihood look-up table decoding.
On Quantinuum's H2-1 device, preparation of the $\ket{\overline{0}}$ state in the [[23,1,7]] Golay code achieves a logical SPAM error rate of $3.3_{-2.4}^{+8.6} \times 10^{-4}$ with an acceptance rate of $47.23(86)\%$.
This estimation surpasses (within $95\%$ confidence intervals) the minimum SPAM error rate of $6.0(1.6) \times 10^{-4}$ for a physical $\ket{0}$, as well as the best previously demonstrated logical state preparations.
\end{abstract}

\maketitle

\section{Introduction} \label{sec:intro}
The reliable preparation of logical stabilizer states is a fundamental requirement for universal fault-tolerant (FT) quantum computation~\cite{Sho95:deco, got02:QEC, nie10:QEC}. 
Logical states such as $\ket{\overline{0}}$ and $\ket{\overline{+}}$ typically serve as initial or resource states in most quantum algorithms~\cite{nie10:QEC}. 
Moreover, future quantum computers with limited connectivity between logical qubit blocks will likely rely on entangled logical stabilizer states to implement logical gates between distant blocks~\cite{Lit19:gamesurf, Thom24:qcompcolor, Yoder25:tourgross}. 
For quantum error-correcting (QEC) codes that encode multiple logical qubits, such as qLDPC codes~\cite{Breu21:qldpc_codes}, diverse logical stabilizer states must be prepared fault-tolerantly and consumed to implement the universal FT Clifford group via teleportation protocols~\cite{Gott99:telpgates, Brun15:telep, Goto24:hypercubes}. 
Furthermore, logical computational qubits may be corrected using Steane~\cite{Ste97:steaneQEC, Brown21:betweenShor, Post24:steaneQECexp} or Knill~\cite{knill04:knillQEC} QEC schemes, both of which require additional logical stabilizer states.

The challenge is to construct FT and resource-efficient preparation circuits for important QEC codes, or families of codes, at any code distance $d$.
The circuit’s fault tolerance ensures that up to $t = \lfloor d/2 \rfloor$ faults, caused by ambient noise or experimental imperfections, are kept under control.
If faults arise at a physical error rate $p$, the FT property guarantees that the probability of a logical error, such as mistakenly preparing $\ket{\overline{1}}$ instead of $\ket{\overline{0}}$, decreases exponentially as $\mathcal{O}(p^{t+1})$~\cite{Pre98:SteaneFT, got02:QEC}, where $d$ is the code distance.
When restricted to Calderbank-Shor-Steane (CSS) codes~\cite{CalShor96:CSS, Ste96:CSS}, the challenge simplifies, as Pauli $X$ and $Z$ faults can be treated independently.

Reduced instances of the non-FT part can be obtained by solving a partial Latin rectangle~\cite{Ste04:fastFTfilter, sal07:CSSEncLatinRect} on the stabilizer generators or via Clifford synthesis methods~\cite{Patel03:clifsyn, Duncan20:clifsym, Webster25:clifsyn, AIStabPrep26:DohPuvBre}.
The verification circuit can be constructed by preparing and consuming several non-FT logical copies of $\ket{\overline{0}}$ and $\ket{\overline{+}}$ to verify one of them either deterministically~\cite{Ste03:filter, Ste04:fastFTfilter} or probabilistically~\cite{Rei04:improvedancprep, Cro09:SteaneFT2, Yi18:largeblockprep}.
While this scheme can be resource-intensive, for particular CSS codes the overhead can be reduced and the need for repetition eliminated~\cite{Pae11:GolayCode}.
More recently, resource-efficient instances of both parts have been discovered by simple inspection of problematic faults in small codes such as the Steane code~\cite{Got16:SteaneCodeFT} and the distance-3 rotated surface code~\cite{got23:sur3}.
For slightly larger codes, including the distance-5 rotated surface code and the distance-5 and -7 color codes, discoveries have been achieved through reinforcement learning~\cite{zen24:reinf.learn} and SAT solvers~\cite{Peh25:automatedSynthesis}.
However, the heavy computational cost of these methods has so far prevented the discovery of resource-efficient FT circuits for larger code distances.

\begin{table*}[t]
    \centering
    \setlength{\tabcolsep}{3pt}
    \small
    \begin{tabular}{L{3.9cm} C{1.3cm} | C{1.3cm} C{1.3cm} C{1.1cm} C{1.1cm} L{2.6cm} C{3.4cm} }
     \toprule
   \textbf{QEC code and logical state prepared} & \textbf{Baseline CXs} & \textbf{CXs} & \textbf{Sim. Qubits} & $\textbf{Flags}$ & $\textbf{Depth}$ & $\textbf{Log. Error Rate}$& $\textbf{Acceptance Rate}$ \\
   \midrule
    $[[7,1,3]]$ Steane $\ket{\overline{0}}$ & $\textbf{11}$ \cite{Got16:SteaneCodeFT} & $15$ & $8$ & $3$ & $10$ & $[2.7 , \,\, 2.9] \times 10^{-5}$ & $[0.9783, \,\, 0.9784]$\\
        \addlinespace[0.8mm]
    $[[9,1,3]]$ rot. surface $\ket{\overline{0}}$ & $\textbf{8}$ \cite{got23:sur3} & $26$ & $12$ & $9$ & $9$ & $[2.4, \,\, 2.6] \times 10^{-5}$ & $[0.97150,\,\,0.97158] $\\
        \addlinespace[0.8mm]
        $[[17,1,5]]$ color code $\ket{\overline{0}}$ & $\textbf{71}$ \cite{Peh25:automatedSynthesis} & $74$ & $23$ & $21$ & $25$ & $[7.7 , \,\, 18.2] \times 10^{-7}$ & $[0.8945,\,\,0.8948]$\\
        \addlinespace[0.8mm]
        $[[25,1,5]]$ rot. surface $\ket{\overline{0}}$ & $120$ \cite{Den02:Topological} & $\textbf{92}$ & $32$ & $28$ & $23$ & $[6.7 , \,\, 24.2] \times 10^{-7}$ & $[0.8980,\,\,0.8984]$\\
        \addlinespace[0.8mm]
        $[[49,1,5]]$ triorthogonal $\ket{\overline{+}}$ & $936$ & $\textbf{361}$ & $95$ & $105$ & $59$ & $[4.2 , \,\, 4.7]\times 10^{-5}$ & $[0.585,\,\,0.584]$\\
        \addlinespace[0.8mm]
        $[[20,2,6]]$ self-dual $\ket{\overline{00}}$& $376$ & $\textbf{145}$ & $36$ & $47$ & $54$ & $[2.3, \,\, 9.7] \times 10^{-8}$ & $[0.8234,\,\,0.8235$]\\
        \addlinespace[0.8mm]
        $[[23,1,7]]$ Golay $\ket{\overline{0}}$ & $297$ \cite{Pae11:GolayCode} & $\textbf{237}$ & $44$ & $80$ & $33$ & $[1.8, \,\, 3.1] \times 10^{-7}$ & $[0.7095,\,\,0.7099]$\\
        \addlinespace[0.8mm]
        $[[31,1,7]]$ color code $\ket{\overline{0}}$ & $421$  \cite{Peh25:automatedSynthesis} & $\textbf{211}$ & $55$ & $69$ & $58$ & $[2.1, \,\, 5.4] \times 10^{-7}$ & $[0.750,\,\,0.751]$\\
        \addlinespace[0.8mm]
        $[[49,1,7]]$ rot. surface $\ket{\overline{0}}$ & $336$  \cite{Den02:Topological} & $\textbf{262}$ & $64$ & $85$ & $46$ & $[1.2 , \,\, 4.4]\times 10^{-7}$ & $[0.702,\,\,0.703]$\\
        \addlinespace[0.8mm]
        $[[95,1,7]]$ triorthogonal $\ket{\overline{+}}$ & $ 4792 $ & $\textbf{1175}$ & $258$ & $380$ & $389$ & $ [4.4, \,\, 6.3] \times 10^{-5} $ & $ [0.240, \,\, 0.241] $\\
        \addlinespace[0.8mm]
        $[[49,1,9]]$ color code $\ket{\overline{0}}$ & $1020$ & $\textbf{408}$ & $93$ & $136$ & $123$ & $[1.1, \,\, 5.8]\times 10^{-7} $ & $[0.531,\,\,0.532]$\\
        \addlinespace[0.8mm]
        $[[81,1,9]]$ rot. surface $\ket{\overline{0}}$ & $720$  \cite{Den02:Topological} & $\textbf{614}$ & $141$ & $206$ & $129$ & $[2.0, \,\, 11] \times 10^{-7}$ & $[0.355,\,\,0.356]$\\
        \addlinespace[0.8mm]
        $[[47,1,11]]$ self-dual $\ket{\overline{0}}$ & $4140$ & $\textbf{1033}$ & $186$ & $388$ & $292$ & $[3.6 , \,\, 17] \times 10^{-7}$ & $[0.122, \,\, 0.123]$\\
        \addlinespace[0.8mm]
        $[[71,1,11]]$ color code $\ket{\overline{0}}$ & $1860$ & $\textbf{829}$ & $177$ & $268$ & $282$ & $[4.4 , \,\, 29] \times 10^{-8}$ & $[0.214, \,\, 0.215]$\\
        \midrule 
     \end{tabular}
     \caption{
     Size and performance of the fault-tolerant (FT) preparation circuits obtained with the \textit{flag-at-origin} construction.
     From left to right, the first two columns indicate: the code and the initial logical state prepared, the number of CX gates required by the best of the baseline and known optimized constructions. 
     Then, for our constructions: the maximum number of simultaneous qubits required, the number of flag qubits, the depth in terms of two-qubit gates, the logical error rate, and the acceptance rate in circuit-level simulations (including memory noise) at a physical error rate of $10^{-3}$, using Wilson confidence intervals of $95\%$. 
     For the even-distance $[[20,2,6]]$ code, the acceptance rate at decoding is $[0.999989, 0.999990]$.}
     \label{tab:circs}
\end{table*}

A baseline alternative construction for CSS codes consists of initializing all code qubits in $\ket{0}$ and fault-tolerantly measuring all the $X$-type stabilizer generators~\cite{nie10:QEC}.
A FT non-deterministic preparation repeats the measurement $\lfloor d/2 \rfloor + 1$ times and restarts upon the detection of any error.
This construction is conceptually simple, modular, and applicable to any CSS code at any code distance, making it suitable for comparison.

In the surface code, the measurement of stabilizer generators can be fault-tolerantly performed through careful gate scheduling~\cite{Den02:Topological, Tomita14:rotsurfschedule}.
However, for most other CSS codes, a resource-efficient FT measurement of the generators involves appending flag gadgets~\cite{Cha18:FTErrCorrd3} to the ancillary qubit where the measurement outcome is recorded.
Discovering flag gadgets that preserve the fault tolerance of the protocol is a challenging problem. Some general and straightforward constructive algorithms exist~\cite{cha18:FlagErrCorrArbitraryDist, Cha20:FTErrCorrAlld}, but they can yield gadgets that consume excessive flag qubits and gates.
Recently, similar flag-based schemes, tailored for specific codes, have been proposed to enable parallel syndrome extraction of various stabilizer operators~\cite{Rei20:FTErrCorrNoExtraQub}. 
\\

This work, as described in Sec.~\ref{sec:construction}, proposes appending flag gadgets to a bipartite non-FT preparation circuit to achieve fault tolerance.
Instead of optimizing the non-FT part, we use a bipartite CX circuit~\cite{Van04:graphicalgraph} to prepare the state, where CX gates control qubits in one partition (\textit{control qubits}) and target qubits in the complementary partition (\textit{target qubits}).
The key observation is that in such bipartite circuits, Pauli-$X$ errors propagate only from control to target qubits, while Pauli-$Z$ errors propagate only from target to control qubits.
Rather than continuing with a verification circuit, flag gadgets are appended at origin: those detecting $X$ errors are appended to control qubits, and those detecting $Z$ errors to target qubits, rendering the circuit FT.
The size of the flag gadget appended to each qubit depends only on the code distance and the number of CX gates acting on that qubit, and is independent of the rest of the code.

Table~\ref{tab:circs} presents the circuit overhead and estimated performance of the proposed construction, showing that for all medium to large codes it outperforms state-of-the-art circuits and baseline constructions.
Even for QEC codes individually optimized, such as the [[31,1,7]] color code or the [[23,1,7]] Golay code, our construction yields reduced circuits, at least in the number of CX gates.
In particular, the proposed construction requires fewer CX gates than the baseline FT preparation for the rotated surface code~\cite{Den02:Topological, Tomita14:rotsurfschedule}.

\begin{figure*}
    \centering
    \includegraphics[width=1\linewidth]{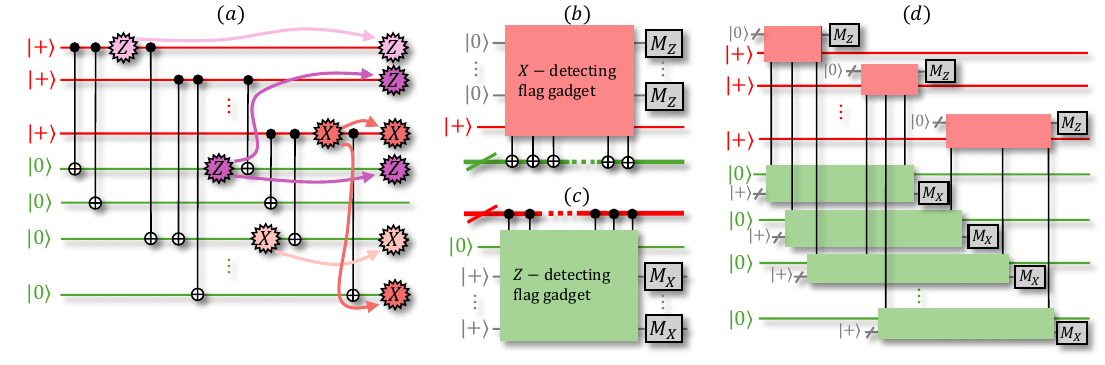}
    \caption{\textit{Flag-at-origin} construction of a \textit{CSS state}, i.e., a stabilizer state whose stabilizer generators are either a tensor product of Pauli-$X$ operators or a tensor product of Pauli-$Z$ operators. $(a)$ Such a state can be prepared non-fault-tolerantly by a bipartite CX circuit, where CX gates control only \textit{control qubits} initialized in $\ket{+}$ and target only \textit{target qubits} initialized in $\ket{0}$. The key observation is that Pauli-$X$ errors propagate only from control to target qubits, while Pauli-$Z$ errors propagate only from target to control qubits. $(b)$ An $X$-detecting flag gadget appended to a control qubit can detect any hook $X$ error occurring on it before it propagates. $(c)$ A $Z$-detecting flag gadget appended to a target qubit can detect any hook $Z$ error occurring on it before it propagates. $(d)$ Appending $X$-detecting flag gadgets to all control qubits and $Z$-detecting flag gadgets to all target qubits results in a fault-tolerant (FT) preparation circuit.}
    \label{fig:construction}
\end{figure*}

Sec.~\ref{sec:flags} describes our algorithm for discovering FT flag gadgets with a minimal number of flag qubits and CX gates.
The algorithm takes as input the code distance, the number of CX gates acting on a qubit, and the user's initial guess for the number of flag qubits, i.e., the gadget size.
Starting from the end of the gadget, the algorithm iteratively attempts to add CX gates until the gadget is found or all possible CX placements are exhausted.
Even without parallelization or advanced high-performance computing tools, optimal FT flag gadgets for tuples (code distance, CX count, number of flag qubits) as large but resource-efficient as (11, 10, 6), (7, 14, 6), or (5, 29, 6) are discovered.
For larger (possibly suboptimal) input numbers of flag qubits, the algorithm produces gadgets of sizes up to (11, 24, $\leq16$) or (7, 71, $\leq25$) within minutes.
We call a gadget ``optimal'' if the algorithm cannot find a FT gadget with one fewer flag qubit, and ``possibly suboptimal'' if the algorithm requires excessive time to check for smaller FT gadgets.
Beyond state preparation, these gadgets are useful for other tasks such as FT syndrome measurements or preparing verified cat states\cite{Shor:97:FTcomp}.

Sec.~\ref{sec:results} presents the numerical and hardware results.
The discovered quantum circuits are benchmarked via subset-sampling Monte Carlo simulations and approximate maximum-likelihood look-up tables, as described in Sec.~\ref{sec:decoding}.
Since the logical SPAM error of the $\ket{\overline{0}}$ state only provides the logical error rate against $X$ errors, additional simulations of a Steane-QEC gadget are performed to obtain performance against $Z$ errors.
Finally, the Golay code FT preparation is demonstrated on Quantinuum's H2-1 device~\cite{H2:specs}.
From the 14,000 preparation attempts submitted, 6,140 were accepted ($47.23(86)\%$ acceptance rate), and among these, only 2 were unsuccessful.
This constitutes an average logical error of $3.3 \times 10^{-4}$ with $95\%$ Wilson confidence intervals of $[0.9, 11.9] \times 10^{-4}$.
This estimation is below (within error bars) the break-even point $6.0(1.6) \times 10^{-4}$ of the SPAM error for a single physical $\ket{0}$ state~\cite{H2:specs} and the state-of-the-art logical error rate demonstrations on hardware~\cite{ryan22:impl, rei24:tess, mayer24:fou, pae24:demons, Dag25:magic_prep}, also on Quantinuum devices.

This \textit{flag-at-origin} construction is general for CSS codes and scalable to large code distances thanks to its simplicity and modularity.
It can construct FT circuits to produce diverse logical states across single or multiple QEC code blocks, such us those employed as a resource to implement addressable and parallel Clifford operations via injection~\cite{Brun15:telep}.
This feature is of particular interest for high-encoding rate CSS codes such as qLDPC~\cite{Breu21:qldpc_codes} or hypercubes~\cite{Goto24:hypercubes} codes.
Unlike other constructions that require solving a global optimization problem for the entire QEC code, the only optimization required here is performed qubit by qubit: discovering a flag gadget for a given code distance and the number of qubits connected to each qubit.
Once discovered, a flag gadget can be reused for the FT preparation of other QEC codes.
As discussed in Sec.~\ref{sec:outlook}, the modularity of the construction leaves room for further optimization, such as the reordering of some commuting gates to reduce circuit volume without compromising fault tolerance, or the combination with existing constructions—like those mentioned previously—to further reduce the CX count.

\section{Fault-tolerant circuit construction} \label{sec:construction}
The section begins with a technical definition of fault tolerance, followed by a description of the proposed construction, guided by Fig.~\ref{fig:construction}.
It then discusses the remaining opportunities for optimizing these circuits and applies them to the Golay code.
Additional optimizations and combinations with other strategies are presented in Sec.~\ref{sec:outlook}.

This work focuses on the FT preparation of CSS states. A \textit{CSS state} is a stabilizer state whose stabilizer generators are each either a tensor product of Pauli-$X$ operators (and $I$) or a tensor product of Pauli-$Z$ operators (and $I$).
Any CSS code $[[n,k,d]]$ (with $n$ physical qubits encoding $k < n$ logical qubits at code distance $d$) prepared in a logical state stabilized by logical operators of this form is a CSS state.
Examples of CSS states are $\ket{\overline{0}}$ or $\ket{\overline{+}}$-the eigenstates of the logical $\overline{Z}$ and $\overline{X}$ logical operators, respectively-in CSS codes with $k=1$, logical GHZ states $(\ket{\overline{0\cdots 0}}+\ket{\overline{+\cdots +}})/\sqrt(2)$ in CSS codes with $k>1$, or any tensor product of them, among others. 
A key advantage of CSS codes is that $X$ and $Z$ errors can be analyzed independently when testing fault tolerance or decoding.

\subsection{Fault tolerance criterion}
Due to hardware imperfections, circuit components can fail with some probability, introducing errors into the quantum circuit.
These faults are typically modeled by replacing the faulty component with its ideal version, followed by the application of a random Pauli error.
Specifically, a random Pauli operator from the set $\{X, Y, Z\}$ is applied after a faulty qubit preparation, single-qubit gate, or qubit idling, and before a qubit measurement.
For a faulty two-qubit gate, a random Pauli operator from $\{I, X, Y, Z\}^{\otimes 2} \setminus \{I\otimes I\}$ is applied.
These errors can propagate through the circuit, potentially resulting in high-weight errors, as illustrated in Fig.~\ref{fig:construction}$(a)$.
To ensure fault tolerance, error propagation must be carefully controlled.

A preparation circuit for a CSS state is FT~\cite{got02:QEC} if, for any number $f \leq t = \lfloor d/2 \rfloor$ of faulty components, the resulting propagated error is either of minimum weight $w \leq f \leq t$ and therefore correctable (or detectable at even distance) by an ideal decoder, or is detected by the ancillas and flags in the circuit, causing the preparation attempt to be restarted.
Therefore, testing fault tolerance requires checking every possible combination of Pauli errors inserted after any set of up to $t$ faulty components.
The minimum weight of a Pauli error is the minimum support of the error under the multiplication by all stabilizer operators of the CSS state, including the logical stabilizers. 
Fortunately, for CSS codes, a circuit satisfies the FT criterion if it holds separately for the cases where only $X$-type faults are inserted and where only $Z$-type faults are inserted, which greatly simplifies verification.

\subsection{\textit{Flag-at-origin} construction}
As shown in Appendix~\ref{app:bipartite}, every CSS state can be prepared by a bipartite CX circuit~\cite{Van04:graphicalgraph}, such as the one in Fig.~\ref{fig:construction}$(a)$, where each qubit serves exclusively as either the control or the target of CX gates.
A key observation is that, in bipartite CX circuits, $Z$ errors on control qubits and $X$ errors on target qubits do not propagate, making them correctable by default and automatically compliant with the fault tolerance criterion for CSS state preparation.
The only propagating errors that must be detected are $X$ errors on control qubits and $Z$ errors on target qubits.

\begin{figure*}
    \centering
    \includegraphics[width=1\linewidth]{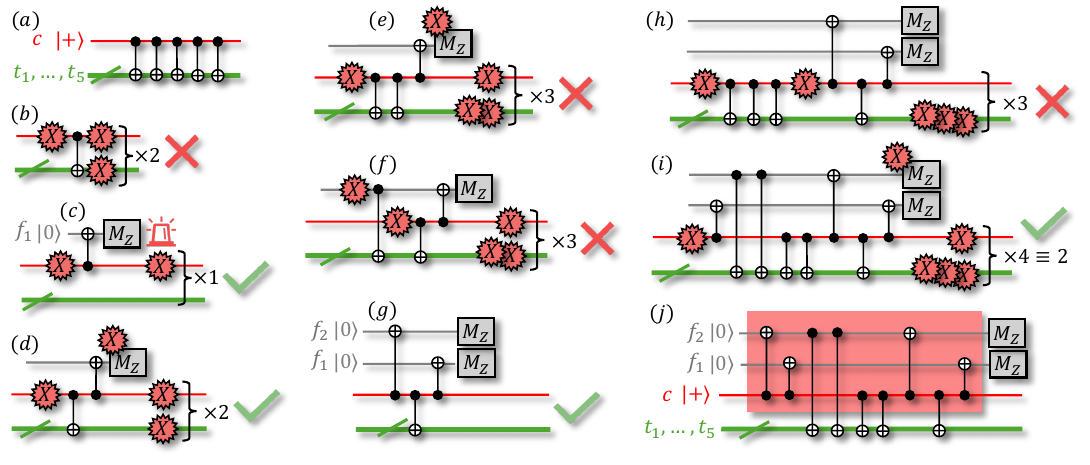}
    \caption{
Algorithm to discover flag gadgets.
$(a)$ The inputs are the number $t = \lfloor d/2 \rfloor$ (equal to 2 in this figure), representing the number of correctable faults for a distance-$d$ code; the number of target qubits (5 in this figure); and the number of flags believed to be sufficient (2 in this figure).
The algorithm constructs the gadget from right to left by iteratively adding new gates from a pool of available gates and testing for fault tolerance.
$(b)$ The first attempted gate does not achieve fault tolerance because a single fault can propagate into two faults.
$(c)$ Adding flag qubit $f_1$ makes the gadget FT (fault-tolerant) because no single fault propagates to an undetected error of weight greater than 1.
$(d)$ Re-adding the first attempted gate now preserves fault tolerance because even two faults (including a faulty measurement) do not propagate to an undetected error of weight greater than 2.
$(e)$ However, adding the next CX gate from $c$ to $t_2$ is not FT because two faults propagate to an undetected weight-3 error.
$(f)$ The next CX in the pool controls $f_1$ instead. This is possible because at this point $f_1$ shares a GHZ-like entanglement with the control qubit, but this attempt is still not FT.
$(g)$ Adding a new flag $f_2$ makes the gadget FT.
$(h)$ After several successful steps, adding the last CX from $c$ to $t_4$ is not FT because two faults propagate undetected to a weight-3 error.
$(i)$ Moving the control of the previous attempted gate to $f_2$ and disentangling $f_1$ makes the gadget FT despite two faults propagating to weight 4. This is because this error reduces to weight 2 up to the weight-6 stabilizer operator created by the circuit for the CSS code.
$(j)$ The algorithm outputs the FT flag gadget.}
    \label{fig:flags}
\end{figure*}

For any input CSS state, the Python library StabGraph~\cite{Amaro19:stabgraph} can generate bipartite CX circuits with a reduced CX count.
Both types of propagating errors are detected at their origin by appending $X$-detecting flag gadgets, as in Fig.~\ref{fig:construction}$(b)$, to every control qubit, and $Z$-detecting flag gadgets, as in Fig.~\ref{fig:construction}$(c)$, to every target qubit.
This yields the FT circuit shown in Fig.~\ref{fig:construction}$(d)$.
The two gadget types must be carefully coordinated to preserve the internal gate order within each gadget, as this ordering guarantees the gadget’s fault tolerance.
An example of an $X$-detecting flag gadget for five CX gates and code distances 4 or 5 is shown in Fig.~\ref{fig:flags}$(j)$.
To maintain fault tolerance, the preparation attempt is restarted whenever any flag gadget detects an error.

In contrast to approaches~\cite{zen24:reinf.learn, Peh25:automatedSynthesis} that optimize the non-FT encoding circuit to minimize the verification overhead, our method employs a larger—but bipartite—encoding circuit, which greatly simplifies the process of achieving fault tolerance.
The entire construction requires only polynomially many resources: the maximum number of gates in a bipartite circuit scales as $\Omega(n^2)$, and, in our observations, the flag gadgets require relatively few CX gates and flag qubits, with sizes growing only linearly in the code distance and in the number of gates acting on each code qubit.
Table~\ref{tab:circs} compares the CX gate count of our construction against state-of-the-art circuits from the literature and against a baseline construction.
The table also reports the maximum number of qubits used simultaneously during circuit execution (assuming fast qubit reset), the circuit depth, and the logical error and acceptance rates obtained under circuit-level simulations with depolarizing noise on operations and memory noise on idle qubits.

\subsection{Room for optimization}
The proposed construction leaves several degrees of freedom that can be exploited to reduce the number of gates and flag qubits, or at least to decrease the circuit depth and/or the maximum number of simultaneously active qubits.
Rather than applying a sophisticated optimization procedure, we explore thousands of random configurations of these degrees of freedom and select the best resulting circuit.
One such degree of freedom, optimized for each code, is the choice of bipartite circuit representation for a given CSS code—a feature already built into StabGraph~\cite{Amaro19:stabgraph}.

Two additional optimizations are applied to the Golay code circuit to reduce the number of flag qubits, gates, and the maximum number of simultaneously active qubits.
The first optimization follows from the observation that, in the Golay code, every $Z$ ($X$) error has maximum weight 3 up to multiplication by a representative of the logical $Z$ ($X$) operator.
An analogous property holds for the Steane code (maximum weight 1) and the $[[47,1,11]]$ code (maximum weight 5).
As a consequence, when preparing the logical $\ket{\overline{0}}$ for the Golay code, the appended flag gadgets do not need to detect combinations of three $Z$-type faults, since these can propagate to at most a weight-3 error up to stabilizer multiplication.
This allows the use of smaller $Z$-detecting flag gadgets designed for distances 4 and 5, rather than for distances 6 and 7, reducing both the number of flag qubits and the CX count.

The second optimization, applicable to all studied codes, leverages the freedom to reorder commuting CX gates in bipartite circuits, before appending the flag gadgets. 
While any such ordering remains FT once the flag gadgets are appended, different orders lead to varying circuit depths and maximum numbers of simultaneous qubits. 
By shuffling these CX gates multiple times and selecting the order that minimizes the maximum number of simultaneous qubits, we obtain a preparation circuit that fits within the 56-qubit limit of the Quantinuum H2-1 device. 
Alternatively, this optimization can be used to reduce the circuit depth. 
Further optimizations are discussed in Sec.~\ref{sec:outlook}.

\section{Discovery of flag gadgets} \label{sec:flags}
This section explains the algorithm proposed to discover flag gadgets.  
The purpose of the gadget is to entangle the target qubits to the control qubit with a FT circuit, i.e., such that any combination of $f \leq t$ faults propagates to a weight $w \leq f$ error up to multiplication with stabilizer operators.  
The explanation is guided by the example depicted in Fig.~\ref{fig:flags}, that describes algorithmic steps leading to the discovery of an $X$-detecting flag gadget with two flag qubits $f_1, f_2$ that protects a control qubit $c$ connected to five target qubits $t_1, \ldots, t_5$ against up to $t = 2$ faults of $X$-type.  
Analogously, to protect a target qubit connected to five control qubits against $Z$-type faults, the $Z$-detecting flag gadget is simply a Hadamard-conjugated version of the gadget in the figure.  
Recall that in a CSS code, $X$-type and $Z$-type errors can be treated independently.  
Pseudo-code describing the algorithm more precisely is provided in Appendix~\ref{app:pseudocode}.  
Table~\ref{tab:flags} shows the number of flag qubits consumed by the flag gadgets discovered.  
These gadgets are of interest beyond this work: for flag-based syndrome extraction, FT preparation of cat states, etc.

Some of the CX gates in the gadget entangle the target qubits to the control qubit. 
The rest entangle and disentangle flag qubits to allow the detection of potentially non-correctable fault combinations, i.e., $f \leq t$ faults propagating to a weight $w > f$ error. 
A measurement output of $-1$ on any of the flags indicates the presence of some potentially non-correctable error, so, when observed, the preparation attempt is discarded and restarted. 
One challenge in the discovery of flag gadgets is that the components that form the gadget are themselves subject to failure.

The algorithm takes as input the number $t$ of correctable faults, the number of target qubits, and an estimate of the number of flag qubits required. 
The flag gadget is then constructed gate by gate, starting from the end of the circuit. 
This approach is necessary because, at each iteration, the fault tolerance of the gadget must be tested, and such testing is only possible once the circuit’s end is defined. 
For clarity, we consider the arrow of time to move from right to left in the circuit. 
Thus, we say, for example, that initially all qubits are ``disentangled'', that the first CX gate from $c$ to $f_1$ in Fig.~\ref{fig:flags}$(c)$ ``entangles'' $f_1$ to $c$, and that applying the same gate later, as in Fig.~\ref{fig:flags}$(i)$, ``disentangles'' $f_1$.

The algorithm must complete three tasks, in the following order of priority, to finalize the gadget: (1) entangle the target qubits to the control qubit, (2) entangle the flag qubits to the control qubit, and (3) disentangle the flag qubits. 
Three pools of CX gates are initialized for each of these tasks and updated as the gadget construction progresses. 
The first pool, responsible for entangling target qubits, is initialized to entangle the first target qubit: $\mathcal{T}_0 = [CX(c, t_1)]$. 
The second pool, responsible for entangling flag qubits, is initialized to entangle the first flag qubit: $\mathcal{E}_0 = [CX(c, f_1)]$. 
The third pool, used to disentangle flag qubits, is initialized as an empty list $\mathcal{D}_0 = \varnothing$, since no flags are entangled at the start. 
Additionally, the set of entangled target qubits is initialized as $T_0 = \varnothing$, and the set of entangled flags as $E_0 = \varnothing$. 
Finally, the output circuit is initialized as an empty list $\mathcal{C}_0 = \varnothing$.

\begin{table*}[t]
    \centering
    \normalsize
    \setlength{\tabcolsep}{3pt}
    \begin{tabular}{l | *{19}{c}}
    \toprule
    tar. qbts.& $3 \!\! - \!\! 4$ & $5$ & $6$ & $7 \!\! - \!\! 8$ & $9 \!\! - \!\! 10$ & $11$ & $12 \!\! - \!\! 13$ & $14$ & $15 \!\! - \!\! 16$ & $17$ & $18$ & $19$ & $20 \!\! - \!\! 21$ & $22$ & $23$ & $24$ & $25$ & $26 \!\! - \!\! 27$ & $28 \!\! - \!\! 29 $ \\
    \midrule
    $t = 1$     & 1 & 1 & 1 & 1 & 1 & 1 & 1 & 1 & 1 & 1 & 1 & 1 & 1 & 1 & 1 & 1 & 1 & 1 & 1 \\
    $t = 2$     & 1 & 2 & 3 & 3 & 3 & 3 & 4 & 4 & 4 & 4 & 5 & 5 & 5 & 5 & 5 & 6 & 6 & 6 & 6 \\
    $t = 3$     & 1 & 2 & 3 & 4 & 5 & 5 & 6 & 6 & $\leq \!\! 7$ & $\leq \!\! 7$ & $\leq \!\! 8$ & $\leq \!\! 8$ & $\leq \!\! 9$ & $\leq \!\! 9$ & $\leq \!\! 9$ & $\leq \!\! 10$ & $\leq \!\! 10$ & $\leq \!\! 11$ & $\leq \!\! 11$ \\
    $t = 4$     & 1 & 2 & 3 & 4 & 6 & 7 & 7 & $\leq \!\! 8$ & $\leq \!\! 9$ & $\leq \!\! 10$ & $\leq \!\! 10$ & $\leq \!\! 11$ & $\leq \!\! 11$ & $\leq \!\! 12$ & $\leq \!\! 13$ & $\leq \!\! 13$ & $\leq \!\! 14$ & $\leq \!\! 14$ & $\leq \!\! 15$  \\
    $t = 5$   & 1 & 2 & 3 & 4 & 6 & $\leq \!\! 9$ & $\leq \!\! 9$ & $\leq \!\! 10$ & $\leq \!\! 11$ & $\leq \!\! 13$ & $\leq \!\! 13$ & $\leq \!\! 13$ & $\leq \!\! 14$ & $\leq \!\! 15$ & $\leq \!\! 15$ & $\leq \!\! 16$ & $-$ & $-$ & $-$ \\
    \bottomrule
    \end{tabular}
    \caption{
    Size of fault-tolerant (FT) flag gadgets discovered to protect one control qubit connected to several target qubits via CX gates at distance $d$, i.e., protected against up to $t=\lfloor d/2 \rfloor$ faults. 
    Column headers indicate the number of target qubits.
    The gadget requires two CX gates for every flag qubit.
    The symbol $\leq$ indicates that the number of flags may not be optimal.
    Additionally, we discovered FT flag gadgets for columns 31 and 71 for row $t=3$ for the baseline construction of the [[95,1,7]] code, which use 12 and 25 flags, respectively.
    For a number of target qubits equal to one or two, no flag qubits are required.}
    \label{tab:flags}
\end{table*}

As shown in Fig.~\ref{fig:flags}$(b)$, the algorithm begins with the first task by adding the initial gate from $\mathcal{T}_0$ to a temporary circuit for which fault tolerance is tested: $\mathcal{C}_\mathrm{temp} = [CX(c, t_1)]$. 
However, the FT test fails because a single fault propagates into a weight-2 error. 
The algorithm then removes the gate from its original pool, updating $\mathcal{T}_1 = \varnothing$, and leaves the other lists unchanged: $\mathcal{E}_1 = \mathcal{E}_0$, $\mathcal{D}_1 = \mathcal{D}_0$, $T_1 = T_0$, $E_1 = E_0$, and $\mathcal{C}_1 = \mathcal{C}_0$.

Since there are no more gates in the first pool, the algorithm proceeds to the second task by adding the gate in $\mathcal{E}_1$ to the temporary circuit, $\mathcal{C}_\mathrm{temp} = [CX(c,f_1)]$, as shown in Fig.~\ref{fig:flags}$(c)$. 
This time, the FT test is passed. 
After this step, $f_1$ shares a GHZ-like entanglement with $c$, meaning it can be used interchangeably with $c$. 
The pools are updated accordingly: $\mathcal{T}_2 = [CX(c,t_1),\, CX(f_1,t_1)]$ to attempt entangling $t_1$ again, $\mathcal{E}_2 = [CX(c,f_2),\, CX(f_1,f_2)]$ to entangle $f_2$, and $\mathcal{D}_2 = [CX(c,f_1),\, CX(f_1,c)]$ to disentangle $f_1$ from $c$ or $c$ from $f_1$. 
The sets of qubits become $T_2 = T_1$, $E_2 = \{f_1\}$, and the circuit is updated with the new gate, $\mathcal{C}_2 = \mathcal{C}_\mathrm{temp}$.

If the algorithm chooses the option of disentangling $c$ from $f_1$, the incoming wire for the control qubit of the code would correspond to $f_1$, and the gadget would teleport it to the output wire $c$ during execution. 
In this case, $c$ and $f_1$ would exchange roles at input, requiring initialization in $\ket{0}$ and $\ket{+}$, respectively. 
However, in all discovered flag gadgets, the algorithm did not choose this option.

Since the first pool is non-empty, the algorithm again prioritizes the first task by adding the first gate from $\mathcal{T}_2$ to the temporary circuit, $\mathcal{C}_\mathrm{temp} = [CX(c,f_1),\, CX(c,t_1)]$. 
As shown in Fig.~\ref{fig:flags}$(d)$, the earlier problematic fault is no longer an issue; even when combined with a measurement fault, the weight of the undetected propagated error remains less than or equal to the number of faults. 
The pools and sets are then updated to $\mathcal{T}_3 = [CX(c,t_2),\, CX(f_1,t_2)]$, $\mathcal{E}_3 = \mathcal{E}_2$, $\mathcal{D}_3 = \mathcal{D}_2$, $T_3 = \{t_1\}$, $E_3 = E_2$, and the circuit becomes $\mathcal{C}_3 = \mathcal{C}_\mathrm{temp}$.
Figs.~\ref{fig:construction}$(e\text{-}i)$ illustrate additional successful and unsuccessful steps. 

Two remarks are in order regarding Fig.~\ref{fig:construction}$(i)$. 
First, target qubits $t_4$ and $t_5$ are entangled via CX gates whose control is a flag qubit ($f_2$) rather than the control qubit $c$. 
This is possible because $c$ and $f_2$ become indistinguishable once they are entangled. 
In fact, since flag qubits can take the role of the control qubit in an $X$-detecting flag gadget—and, analogously, the role of the target qubit in a $Z$-detecting flag gadget—it is possible for the final preparation circuit to contain a CX gate that controls a flag from an $X$-detecting gadget and targets a flag from a $Z$-detecting gadget. 
Second, the particular two-fault combination shown propagates to an undetected weight-4 error. 
However, the stabilizer operator $X_cX_{t_1}X_{t_2}X_{t_3}X_{t_4}X_{t_5}$—obtained by propagating the stabilizer $X_c$ of the control qubit’s initial state $\ket{+}$ through the circuit—reduces this error to weight 2 under multiplication. 
This reduction ensures that the error is correctable and helps the gadget pass the FT test during the algorithm’s execution. 
Multiplication by other stabilizer operators of the CSS state could further reduce the error weight, potentially leading to shorter execution times and smaller gadgets, though at the cost of possibly compromising the modularity of the construction.

The algorithm terminates when the exit criterion is met, i.e., $T_s$ contains all target qubits, $E_s$ is empty for some step $s$, and the fault tolerance test is passed.  
In the example shown, this occurs with the gadget in Fig.~\ref{fig:flags}$(j)$. 
If, at any step $s$ in the iterative process, all three pools are empty before the exit criterion is satisfied, the algorithm backtracks to the previous successful step, selects the next unused gate from the corresponding pools, and continues constructing a new branch from that point. 
If all gates in all previous steps are exhausted without meeting the exit criterion, the algorithm halts with no FT gadget found for the given inputs.
By incrementally increasing the number of flag qubits provided as input, this procedure can be used to discover optimal gadgets, since the first successful solution found corresponds to the minimal number of flags needed.

However, for some large input values of $t$ (number of correctable faults), a large number of target qubits, and a small number of flag qubits, no gadget can be found within a reasonable time. In such cases, we provide a larger number of flag qubits than initially expected, enabling the algorithm to discover a solution within seconds or minutes. 
These suboptimal flag counts are indicated in Table~\ref{tab:flags} with the $\leq$ symbol. 
For distances larger than $d=11$ ($t=5$), our current FT test is too slow to find a solution, even when the number of flags is not optimal. 
We nevertheless expect that a faster error-propagation routine and algorithm parallelization would allow us to extend the method to higher distances.

\section{Results} \label{sec:results}
This section presents the numerical and hardware results, while the decoder details are given in Sec.~\ref{sec:decoding}. 
State-preparation circuits are first constructed for the codes listed in Table~\ref{tab:circs}, and their sizes are compared against both the state-of-the-art and baseline constructions. 
The fault tolerance criteria are exhaustively tested for the [[7,1,3]], [[17,1,5]], [[20,2,6]], and [[23,1,7]] codes. 
For the triorthogonal codes, the more relevant logical $\ket{\overline{+}}$ state is prepared instead of $\ket{\overline{0}}$, as the transversal $T$~gate acts non-trivially on it.
Therefore, all protocols for triorthogonal codes studied in this work employ the Hadamard-conjugated versions of the states, operators, and Pauli errors.

The noise model employed in the numerical estimation of acceptance and logical error rates is then described. 
Table~\ref{tab:circs} also reports these numerical estimates for all codes at a physical error rate of $p=10^{-3}$, close to present-day hardware error rates. 
For the smallest codes, acceptance and logical error rates are plotted in the range $p \in [0.5, 10] \times 10^{-3}$, confirming that the scaling of logical error rates is consistent with fault tolerance up to the full distances of the codes.
Since preparing a logical $\ket{\overline{0}}$ in a CSS code only probes performance against $X$ errors, additional numerical simulations of Steane-QEC gadgets are performed to assess performance against $Z$ errors.

\subsection{Circuit sizes obtained}
As shown in Table~\ref{tab:circs}, our general-purpose construction is outperformed in CX count by hand-crafted circuits for the Steane code and the distance-3 rotated surface code, and by SAT-solver-based constructions for the [[17,1,5]] color code. 

This is to be expected: for small codes, experienced inspection or computationally intensive approaches such as SAT solvers can explore and discard a large portion of the solution space in the search for an optimal circuit. 
However, for all other codes studied in this work—including the still small distance-5 rotated surface code—our general construction achieves lower CX counts.

Additionally, our construction also outperforms the baseline approach on all QEC codes, particularly for the largest ones, where the CX count is reduced by more than a factor of four. 
It is worth noting that the baseline construction we compare against already incorporates our optimized flag gadgets for stabilizer-generator measurements. 
This approach also surpasses the baseline for rotated surface codes, which do not require flags thanks to their particular gate scheduling that preserves fault tolerance~\cite{Den02:Topological, Tomita14:rotsurfschedule}.

\subsection{Noise model} \label{ssec:noise_model}
The noise model employed is standard in QEC numerical experiments, with the additional inclusion of the often-overlooked memory noise channel. 
The entire model is parameterized by a single error rate $p$, which denotes the probability of introducing a non-identity Pauli error at any fault location in the circuit. 
Specifically, each qubit initialization is followed by a single-qubit depolarizing channel with error rate $p$, every two-qubit gate is followed by a two-qubit depolarizing channel with error rate $p$, and every flag qubit measurement is preceded by a single-qubit bit-flip channel with error rate $p$. 
The measurement of code qubits at the end of the circuit is performed noiselessly, as this measurement does not occur immediately after state preparation in real-world scenarios.
Therefore, the logical error rates estimated are not affected by the noise of the destructive measurement of the code qubits. 

\begin{figure}[t]
	\centering
	\resizebox{0.49\textwidth}{!}{\input{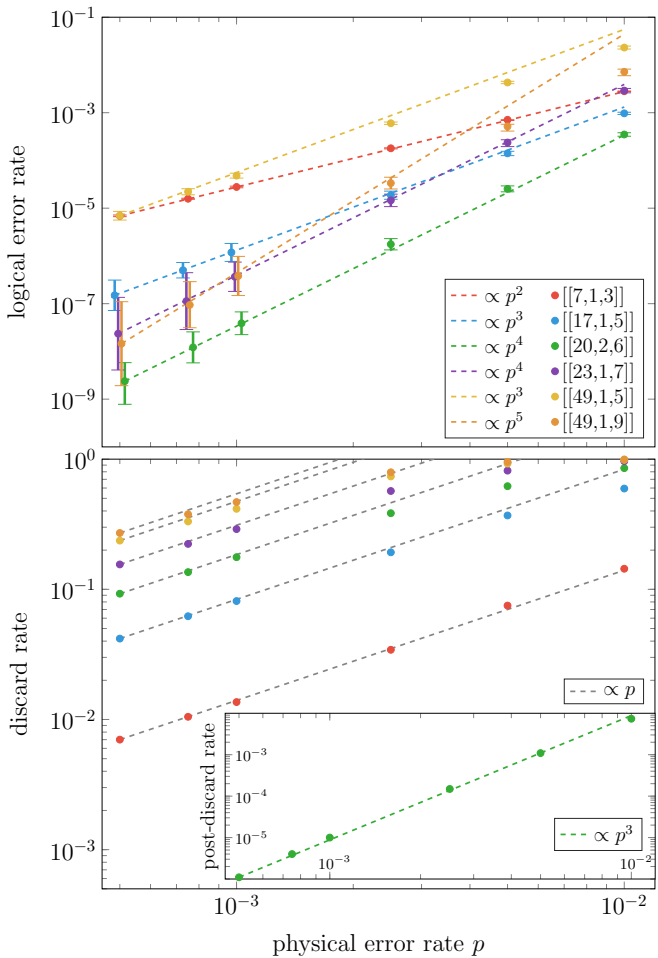} } 
	\caption{
    Numerical logical error rate (top) and discard rate (bottom) as a function of the physical error rate $p$ for various CSS codes. 
    Dashed lines visually indicate the expected $\mathcal{O}(p^{t+1})$ scaling with the code distance $d$, where $t = \lfloor d/2 \rfloor$ is the number of correctable faults. 
    The subplot below shows the post-discard rate of the [[20,2,6]] code during decoding, reflecting its even distance. 
    Error bars indicate $95\%$ confidence intervals.
    }
		\label{fig:log_err}
\end{figure}

Memory noise is introduced by a single-qubit depolarizing channel of error rate $p/100$ on every idle location.
We choose to introduce one idle location for every active qubit for every CX gate, i.e., as if only one CX gate was implemented per unit of time and memory noise accumulated during each of those times on all active qubits.
In order to reduce the memory noise accumulated in the circuit, every qubit is initialized as late as possible, i.e., immediately before the first CX gate acts on it, and every flag qubit is measured as soon as possible, i.e., immediately after the last CX acts on it.
Code qubits are all measured simultaneously after the last flag qubit is measured and not earlier in order to represent the accumulation of memory noise before the next logical operation is performed in a real-life scenario.

\subsection{Performance of preparation circuits}
Table~\ref{tab:circs} shows manageable acceptance rates above $50\%$ for the small-to-medium CSS codes and larger than $20\%$ for the biggest codes of distance $d=11$.
The logical error rates reported are clearly below the break-even point of $p=10^{-3}$ and as low as $\sim 10^{-7}$ to even $\sim 10^{-8}$ for the best-performing codes in the middle of the table.
As expected, the logical error rates decrease with increasing distance, but for the largest codes the logical error rate unexpectedly saturates to the regime of $10^{-5}$.
Sec.~\ref{sec:decoding} discusses how the increase in error rate on the second half of the table is likely due to our decoding pipeline being limited at large code sizes.

Additionally, the discard rate (1 minus acceptance rate) and logical error rate as a function of the physical error rate is plotted in Fig.~\ref{fig:log_err} for the smallest CSS codes.
The figure shows the expected logical error rate decays $\mathcal{O}(p^{t+1})$ as dashed lines for every CSS code.
One can see that for the smallest CSS codes in the regime of low error rates, the expected decay is matched by the numerical results.
This is a numerical evidence that the preparation circuits are indeed FT up to $t$ faults.
For the two largest codes analyzed, the correct scaling might become evident at even lower values of the physical error rate, but that regime is computationally harder to explore.

In the special case of even-distance codes some errors of weight $t=d/2$ can be detected but not corrected.
Therefore, if a syndrome is compatible with a weight-$t$ error but not with lower-weight errors, instead of correcting the state, the entire computation must be discarded during decoding, even after a successful state preparation.
The resulting \textit{post-discard rate} for the [[20,2,6]] code is plotted in the subplot at the bottom of Fig.~\ref{fig:log_err}, sowing the expected trend of $\mathcal{O}(p^t)$.

\subsection{Performance of Steane QEC}
The logical error rate in the preparation of a logical $\ket{\overline{0}}$ state is only affected by $X$ errors.
So, the fault tolerance and performance against $Z$ errors is numerically studied with a Steane-QEC gadget.
Appendix~\ref{app:discussion} discusses and demonstrates the importance of protecting the preparation circuits against both types of errors.

In the Steane-QEC gadget, a logical $\ket{\overline{0}}$ resource state is fault-tolerantly prepared, a transversal CX is implemented from the resource block to the computational block that is meant to be corrected, and finally the resource block is measured destructively in the $X$ basis.
Decoding the output provides information about the joint $Z$ errors produced on the blocks and can be used to apply a suitable correction on the computational block.

In realistic settings, logical operations are expected to introduce more noise than the freshly prepared resource state.
Without this imbalance, performing Steane QEC between the logical gates would only serve to increase the noise level of the computational block.
In order to reproduce the effect of noise in this setting, we design the following experiment: a computational block is noiselessly prepared in the logical $\ket{\overline{+}}$ state, single-qubit depolarizing channels of error rate $10p$ are applied on every qubit of the computational block, Steane QEC is performed, and finally, the same depolarizing channel is applied on the computational qubit again. 
The logical resource state $\ket{\overline{0}}$ for Steane QEC is prepared using our construction and subjected to the noise model defined in Sec.~\ref{ssec:noise_model} with noise rate $p$.
The same noise model of rate $p$ is applied to the subsequent transversal CX gate, to the additional idle locations in the Steane-QEC gadget, and to the measurement of the resource state.
The logical error rate of this experiment is compared to the logical error rate of the same experiment without Steane QEC.

\begin{figure}[t]
	\centering
	\resizebox{0.49\textwidth}{!}{ 
    \input{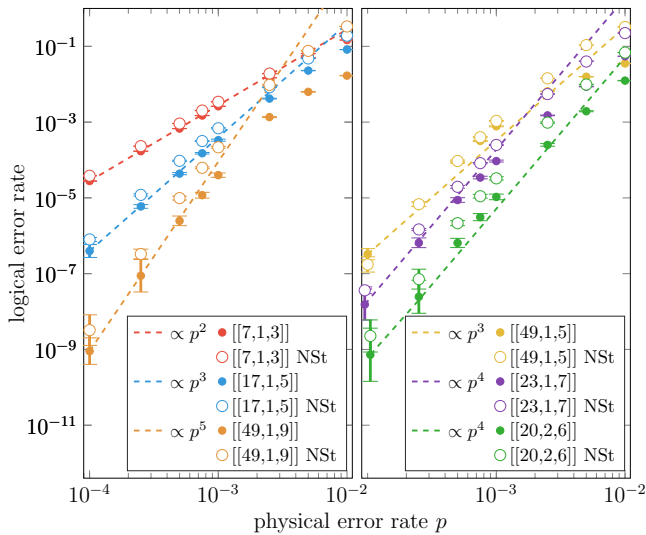} 
	} 
	\caption{
    Numerical logical error rate, with and without (indicated by NSt in the legend) a Steane-QEC gadget, as a function of the physical error rate $p$ for various CSS codes. 
    Dashed lines indicate the expected $\mathcal{O}(p^{t+1})$ scaling with code distance $d$, where $t=\lfloor d/2 \rfloor$ is the number of non-problematic faults. 
    Error bars represent $95\%$ confidence intervals. 
    }
		\label{fig:Steane}
\end{figure}

Figure~\ref{fig:Steane} presents the logical error rates as a function of the physical error rate $p$. 
For five of the six CSS codes studied, performing Steane QEC reduces the logical error rate across the entire simulated range by up to a factor of two. 
The [[49,1,5]] triorthogonal code is an exception, showing no clear improvement over the full range, likely due to the substantial circuit overhead required for FT preparation. 
Expected $\mathcal{O}(p^{t+1})$ scalings are shown as dashed lines for each CSS code; apart from the [[49,1,5]] case, the results follow the predicted scaling in the low-$p$ regime.

\subsection{Hardware results}
Finally, we implement the FT preparation of the $\ket{\overline{0}}$ state for the Golay code on the H2-1 and H2-2 Quantinuum devices.
The results are summarized in Table~\ref{tab:hardware}.
Decoding is performed using look-up tables generated from numerical simulations at an error rate of $p=10^{-3}$, as described in Sec.~\ref{sec:decoding}.
To optimize performance, we explore four different settings for mitigating memory noise, combining compilations in terms of either CX or CZ gates with three variants of dynamical decoupling (DD)~\cite{viola99:dd}.

A portion of memory noise in Quantinuum devices can be approximated by single-qubit coherent rotations $\exp(-i \delta Z)$ on every qubit at every time step, with the same small angle $\delta$\cite{DeCross25:spam}.
If unmitigated, these rotations accumulate over time and propagate to other qubits through CX gates.
Compiling the circuit in terms of CX gates with CZ gates prevents this propagation, since $Z$-type noise commutes with the unwanted rotations.
Another benefit of this compilation is that CZ gates are nearly native in Quantinuum hardware, whereas CX gates require extra single-qubit rotations that can introduce further noise.
To combat the accumulation of coherent rotations, dynamical decoupling (DD) inserts Pauli-$X$ and Pauli-$Y$ operators at regular intervals without affecting the final logical state.
These operators reverse the direction of the coherent rotations, effectively canceling them.
We test three DD configurations:
i) no DD,
ii) the default software DD in Quantinuum devices\cite{DeCross25:spam}, which accounts for precise timing and ion positioning, and
iii) a custom DD that simply inserts an $XX$ Pauli operator after every two-qubit gate.
Similarly to the experimental setting employed in the previous numerical simulations, all qubits are initialized as late as possible, all flag qubits are measured as early as possible, and all code qubits are measured simultaneously after the last measured flag qubit.

All experiments achieve reasonable acceptance rates, ranging from $38\%$ to $51\%$.
The best-performing settings use CZ gates and/or some form of dynamical decoupling (DD).
Aggregating the four experiments E3–E6 on H2-1, which use CZ gates with varying DD schemes, yields 6,140 accepted runs out of 13,000 preparation attempts—an acceptance rate of $47.23(86)\%$.
Among these accepted attempts, only two fail, corresponding to an average logical error rate of $3.3\times 10^{-4}$ with a $95\%$ Wilson confidence interval of $[0.9, 11.9]\times 10^{-4}$.
H2-2 shows a slightly higher average logical error rate of $5.18\times 10^{-4}$ with a confidence interval of $[1.4, 18.9]\times 10^{-4}$.
If syndromes caused by weight-3 errors are discarded rather than corrected (last two rows of Table~\ref{tab:hardware}), experiments E3–E6 produce zero logical errors—with $[0.0, 6.3] \times 10^{-4}$ confidence intervals—across 6,121 post-accepted runs, at the cost of a post-acceptance rate of $99.69^{+0.11}_{-0.17}\%$.

For context, the logical error rate without additional post-discarding is better (within $95\%$ confidence intervals) than the minimum hardware SPAM error rate of $6.0(1.6) \times 10^{-4}$ of a physical $\ket{0}$ in Quantinuum H2-1 and H2-2~\cite{H2:specs}.
It is also consistent with the state-of-the-art logical error rates reported in QEC experiments with distances greater than two and with moderate post-selection: $8^{+16}_{-6} \times 10^{-4}$ with the tesseract code~\cite{rei24:tess}, and with the Steane code, $5.1(2.7) \times 10^{-4}$\cite{Dag25:magic_prep}, $9(2) \times 10^{-4}$~\cite{mayer24:fou}, $5^{+4}_{-3} \times 10^{-4}$~\cite{pae24:demons}, and $4.1(1.3) \times 10^{-4}$~\cite{ryan22:impl}.

The four logical errors observed in experiments E6 and E7 are likely due to the combination of two factors.
First, the maximum-likelihood look-up table used in the first layer of our decoding pipeline does not include the syndromes corresponding to these errors.
Consequently, these cases are passed to the less powerful second layer, which performs minimum-weight decoding.
Second, the relevant syndromes can only be generated by errors of weight at least 3, which can be particularly challenging to correct in a distance-7 code.
The decoding pipeline is described in detail in the next section.

\begin{table}[t]
    \centering
    \setlength{\tabcolsep}{3pt}
    \label{tab:SecondTab}
    \small
    \begin{tabular}{l | C{0.75cm} C{0.75cm} C{0.75cm} C{0.75cm} C{0.75cm} C{0.75cm} C{0.75cm}} 
    \toprule
    & E1 & E2 & E3 & E4 & E5 & E6 & E7\\
    \midrule
        Machine & H2-1 & H2-1 & H2-1 & H2-1 & H2-1 & H2-1 & H2-2 \\
        2q gates & CX & CX & CZ & CZ & CZ & CZ & CZ\\
        \addlinespace[0.8mm]
        DD &None &Def. & None & Def. & Cust. & Def. & Def. \\
        \addlinespace[0.8mm]
        Attempts &$1000$ & $1000$& $1000$ & $1000$ & $1000$ & $10000$ & $10000$ \\
        \addlinespace[0.8mm]
        Acc. Att. & 411 & 509 &389  & 493 & 426 &  4832 & 3859 \\ 
        \addlinespace[0.8mm]
        LE no post. & 10 & 1 & 0 & 0 & 0 & 2 & 2 \\
        \addlinespace[0.8mm]
        Post-acc. & 24 & 5 & 0 & 1 & 2 & 16 & 11 \\
        \addlinespace[0.8mm]
        LE post. & 2 & 0 & 0 & 0 & 0 & 0 & 0 \\
        \midrule 
     \end{tabular}
     \caption{
     Hardware experiment settings and results. 
     In descending order by row: Quantinuum hardware, compilation type (CX or CZ gates), dynamical decoupling (DD) strategy used (none, default, or custom), number of initialization attempts, number of accepted attempts, number of logical errors within them, number post-accepted attempts, and number of logical errors after within them. 
     }
     \label{tab:hardware}
\end{table}

\section{Decoding} \label{sec:decoding}
To estimate the logical error rate of the prepared circuits, the output of the simulated preparation attempts must be decoded.
However, decoding large code distances (up to 11) at low physical error rates (as low as $5 \times 10^{-4}$), with enough precision to resolve logical error rates smaller than $10^{-8}$, poses significant numerical challenges.
To achieve sufficient statistical accuracy, we perform high-speed circuit-level (CL) Pauli propagation simulations, generating up to $10^9$ random Pauli errors drawn from the noise model in Sec.~\ref{ssec:noise_model} for each circuit and physical error rate studied.
This dataset is further expanded up to $10^{10}$ errors via subset-sampling.

Given the computational cost of algorithmic decoders in this regime, we opted for generating maximum-likelihood (ML) look-up tables (LUTs) from the dataset. 
Half of the error set is used as a training set to generate the CL-ML-LUT, while the remaining test set is used as a test set to compute logical error rates.
Since the CL-ML-LUT might not cover all syndromes appearing in the test set, we supplement it with a code-capacity minimum-weight LUT (CC-MW-LUT) to handle previously unseen syndromes.
As an example, generating the CL-ML-LUT for the [[71,1,11]] color code took five days and successfully corrected $99.89\%$ of accepted errors in the test set.
Nonetheless, as discussed later, this pipelined decoding approach struggles to suppress logical error rates below $10^{-5}$ for the largest CSS codes, likely due to limitations of the final decoding layer.

\subsection{Subset-sampling}
In Monte Carlo simulations of Pauli error channels, at moderately low physical error rates $p$, the most likely error is the trivial error $I$, which yields the trivial syndrome (no stabilizer excitations) and no logical error.
To avoid spending computational resources sampling these trivial errors excessively, subset-sampling techniques are employed~\cite{Maur19:subset, Trout18:subset}.

For the two error rates considered in the noise model of Sec.~\ref{ssec:noise_model}, namely $p$ and $q = p/100$, we first count the number of circuit locations $L_p$ and $L_q$ where faults can occur at these respective rates.
From these, we compute the normalized probability distribution $\mathcal{P}_{p,q}(f_p, f_q)$ over the number of faults $f_p$ and $f_q$ at those locations.
By excluding the trivial case of no faults, we obtain a normalized distribution $\mathcal{Q}_{p,q}(f_p, f_q)$ with $\mathcal{Q}_{p,q}(0,0) = 0$.
To limit memory usage, extremely unlikely fault events are excluded by removing $(f_p, f_q)$ pairs for which $\mathcal{P}_{p,q}(f_p, f_q) \leq 1/S^2$, where $S$ is the number of Monte Carlo samples.
We then draw $S$ fault events from $\mathcal{Q}$ and, for each, randomly generate Pauli errors according to the noise model. 
The expected number of missing trivial fault events is added back to the propagated error set by saving their known trivial syndrome, effectively increasing the total number of samples.
For example, in our numerical experiments for the Steane code at $p=10^{-2}$, the error set size grows from $S=10^7$ to over $3.4 \times 10^7$ samples, while for the Golay code at $p=5 \times 10^{-4}$, the set increases from $10^8$ to more than $76 \times 10^8$ samples.

\subsection{Circuit-level maximum-likelihood look-up table}
To generate the circuit-level maximum likelihood look-up table (CL-ML-LUT), non-trivial errors in the final error set are propagated through the circuit using Clifford simulation. 
During this propagation, errors that trigger a flag measurement are counted to estimate the acceptance rate.
For all other errors—including trivial ones—their syndromes and equivalence classes are recorded. 
The equivalence class is determined by whether the error commutes or anticommutes with a chosen logical operator representative that stabilizes the logical state: logical $\overline{Z}$ for $\ket{\overline{0}}$ and logical $\overline{X}$ for $\ket{\overline{+}}$.
For example, the [[20,2,6]] code prepares two logical qubits in the logical state $\ket{\overline{00}}$, so there are four equivalence classes depending on whether the error commutes or anticommutes with each of the two logical $\overline{Z}$ operators.
To prevent overfitting, the dataset of (syndrome, equivalence class) pairs is randomly split into two equal subsets: one used as the training set to generate the CL-ML-LUT, and the other as the test set for logical error rate evaluation.
The CL-ML-LUT stores the most likely equivalence class for each syndrome observed in the training set.
An error in the test set is considered uncorrected if its equivalence class differs from the predicted class in the CL-ML-LUT.
If the syndrome for a test error is absent in the CL-ML-LUT, the decoding proceeds to the next layer in the pipeline, a code-capacity minimum-weight look-up table (CC-MW-LUT), to attempt correction.

\subsection{Code-capacity minimum-weight look-up table}
To generate the code-capacity minimum-weight look-up table (CC-MW-LUT) for the preparation of the logical $\ket{\overline{0}}$ state, we systematically consider all $X$ errors of weight from 1 up to $t$ on the ideally-prepared state and record their corresponding (syndrome, equivalence class) pairs.
Since all these errors are correctable by definition, any errors that produce the same syndrome must belong to the same equivalence class. 
Therefore, the CC-MW-LUT can be constructed as a dictionary that maps each syndrome to a unique equivalence class.
For the [[71,1,11]] color code, due to computational and memory constraints, we restrict the enumeration to errors of weight up to 4 instead of $t=5$.

\subsection{Decoding pipeline}
For every pair (syndrome, equivalence class) in the test set, decoding is first attempted using the CL-ML-LUT.
If the syndrome is not found in the CL-ML-LUT, the CC-MW-LUT is consulted next.
In the event that the syndrome is absent from both look-up tables, a final decoding stage based on an algorithmic decoder is applied. 
For the results reported in Table~\ref{tab:circs}, we employ the Tesseract decoder at this stage \cite{BenHigShu25:Tesseract}. 
This additional step ensures that a correction is produced even when the syndrome is not represented in either look-up table. 
Although algorithmic decoders can be more computationally demanding than a direct look-up, only a small minority of shots require this final decoding stage, so the overall decoding procedure remains efficient. 
Alternative strategies for this final decoding layer are discussed and compared in Appendix~\ref{app:decoding}.

For the [[20,2,6]] code, which has even distance, we do not expect to correct errors of weight $t=3$. 
Instead, the entire computation is discarded whenever such errors are detected.
Accordingly, the decoding pipeline is modified to first decide whether the error should be discarded. 
If not discarded, the error is then corrected using the two LUTs as usual.
Specifically, any error causing a syndrome uniquely associated with a weight-$t=3$ error—i.e., not compatible with any lower-weight correctable error—is discarded.
Although this discard step is not strictly necessary for the distance-7 Golay code, the same decoding pipeline was applied to its preparation on Quantinuum hardware (see the last two rows of Table~\ref{tab:hardware}), resulting in fewer logical errors at the cost of a reduced post-acceptance rate.

Let us now provide more detail about the behavior of the decoding pipeline for the [[71,1,11]] color code at a physical error rate of $p=10^{-3}$.
A total of $S = 3 \times 10^{8}$ error samples are drawn from the distribution $\mathcal{Q}$, which corresponds to 327,796,174 samples from the original distribution $\mathcal{P}$ after subset-sampling—an increase of approximately $9.3\%$.
Half of these, 163,898,087 samples, are randomly assigned to the test set—and the rest to the training set for generating the CL-ML-LUT.
Among the error samples in the test set, 35,126,300 errors are accepted (i.e., do not raise any flag), yielding the $21.4\%$ acceptance rate reported in Table~\ref{tab:circs}.
Of the accepted errors, 35,088,809 ($99.89\%$) have syndromes contained in the CL-ML-LUT, and all of these are successfully corrected.
From the remaining accepted errors, 35,563 ($94.9\%$ of that remainder) have syndromes found in the CC-MW-LUT, and all of these are also successfully corrected.
Finally, for the last 1,928 accepted errors whose syndromes are not contained in either LUT, 4 of them result in a logical error.
This leads to an average final logical error rate in the range $[4.4,29] \times 10^{-8}$.

\section{Outlook} \label{sec:outlook}

The simplicity and flexibility of the proposed construction for FT preparation of CSS states motivate various further optimizations.

For instance, the order of the CX gates in the bipartite CX circuit (before appending the flag gadgets) is a degree of freedom that can be optimized to reduce circuit depth or to minimize the maximum number of simultaneously active qubits, all without compromising fault tolerance.
A partial reordering was implemented for the Golay code to reduce the maximum number of simultaneous qubits below the 56-qubit limit of the H2-series devices.

Another promising optimization is to combine our construction with verification-based approaches~\cite{Peh25:automatedSynthesis, zen24:reinf.learn}, where the former focuses, for instance, on detecting $X$ errors and the latter on $Z$ errors.
This tandem approach is mutually beneficial: all flag gadgets can be applied sequentially, which significantly lowers the maximum number of simultaneous qubits required, and the verification overhead is simplified by reducing the number of errors to verify, easing the classical computational cost of discovering verification circuits.

In correlated QEC schemes that jointly decode multiple logical blocks across a portion of the logical circuit~\cite{QuERA25:corrdec, Zhou24:algFT}, the fault tolerance requirements for each individual gadget can be relaxed. 
Our construction can readily adapt to such settings by employing smaller flag gadgets.
Relatedly, the circuit overhead can be further reduced by replacing the traditional fault tolerance criterion—which ensures propagated errors have low total weight~\cite{got02:QEC}—with a more relaxed criterion, akin to relaxed fault tolerance criteria explored in surface codes~\cite{Den02:Topological, Tomita14:rotsurfschedule}.

Regarding the flag gadgets, one promising optimization lies in exploring and utilizing flag gadgets shared by multiple control or target qubits, following the spirit of recent works~\cite{Liou23:parallelflag, Rei21:flagsSteane}.
Moreover, following the publication of this work, efficient methods for constructing large-distance flag gadgets have been developed. These approaches eliminate any potential overhead associated with building flag gadgets at arbitrary distances \cite{SpiderCat26:KheLiPoo, MQT26:PehWeiWil}.
Beyond algorithms for identifying error-detecting flag gadgets, these techniques could be extended to construct error-correcting flag gadgets, which may offer improved resource efficiency compared to existing constructions~\cite{Cha20:FTErrCorrAlld}. Incorporating such error-correcting flag gadgets into our framework would enable a fully deterministic state-preparation protocol with no discard rate, akin to~\cite{Schmid25:deterministic}.
Taken together, these developments ensure that the resulting scheme applies efficiently at any code distance.

As for decoding, the inclusion of a final algorithmic decoding stage ensures that a correction can be produced for syndromes not contained in the look-up tables, a situation that becomes increasingly relevant for larger codes.
Further improvements may nonetheless be obtained by refining the construction of the CL-ML-LUT, for instance by increasing the number $S$ of samples used in the Monte Carlo simulation or by extending the subset sampling technique to better capture less likely error configurations.
Additionally, selectively discarding runs associated with extremely rare syndromes and nearly degenerate equivalence classes could further reduce the logical error rate.

In conclusion, we expect that the encouraging results demonstrated in this work, combined with the avenues for further optimization, will motivate the adoption of the \textit{flag-at-origin} construction as the standard, general approach for FT preparation of diverse CSS states at arbitrary distances.
\\

\noindent\textbf{Acknowledgments:} 
We thank the entire Quantinuum team for their many contributions that made this research possible. Special thanks to Ben Criger, Ali Lavasani, Ciaran Ryan-Anderson, Callum Macpherson, Natalie Brown, Pablo Andres-Martinez, and Silas Dilkes for their valuable suggestions. 

\noindent\textbf{Data availability}
The code definitions, circuits constructed, flag gadgets discovered, numerical and hardware experimental data to generate the look-up tables for decoding, and the processing scripts necessary to reproduce the tables and figures in this work are available at~\cite{Forlivesi:repo}.

\bibliography{references}

@misc{H2:specs,
  title = {H2 Spec Sheet},
  year = 	{2025},
  month =   {June},
  author = {Quantinuum},
  url = {https://docs.quantinuum.com/systems/data_sheets/Quantinuum%20H2%20Product%20Data%20Sheet.pdf},
}

@misc{Forlivesi:repo,
  author = {Forlivesi, Diego},
  title = {flag-at-origin-paper},
  year = {2025},
  month = {June},
  publisher = {GitHub},
  journal = {GitHub repository},
  url = {https://github.com/CQCL/flag_at_origin_paper},
}

@misc{Schmid25:deterministic,
      title={Deterministic Fault-Tolerant State Preparation for Near-Term Quantum Error Correction: Automatic Synthesis Using Boolean Satisfiability}, 
      author={Ludwig Schmid and Tom Peham and Lucas Berent and Markus Müller and Robert Wille},
      year={2025},
      eprint={2501.05527},
      archivePrefix={arXiv},
      primaryClass={quant-ph},
      url={https://arxiv.org/abs/2501.05527}, 
}

@misc{Dag25:magic_prep,
      title={Experimental demonstration of high-fidelity logical magic states from code switching}, 
      author={Lucas Daguerre and Robin Blume-Kohout and Natalie C. Brown and David Hayes and Isaac H. Kim},
      year={2025},
      eprint={2506.14169},
      archivePrefix={arXiv},
      primaryClass={quant-ph},
      url={https://arxiv.org/abs/2506.14169}, 
}

@article{DeCross25:spam,
  title = {Computational Power of Random Quantum Circuits in Arbitrary Geometries},
  author = {DeCross, M. and Haghshenas, R. and Liu, M. and Rinaldi, E. and Gray, J. and Alexeev, Y. and Baldwin, C. H. and Bartolotta, J. P. and Bohn, M. and Chertkov, E. and Cline, J. and Colina, J. and DelVento, D. and Dreiling, J. M. and Foltz, C. and Gaebler, J. P. and Gatterman, T. M. and Gilbreth, C. N. and Giles, J. and Gresh, D. and Hall, A. and Hankin, A. and Hansen, A. and Hewitt, N. and Hoffman, I. and Holliman, C. and Hutson, R. B. and Jacobs, T. and Johansen, J. and Lee, P. J. and Lehman, E. and Lucchetti, D. and Lykov, D. and Madjarov, I. S. and Mathewson, B. and Mayer, K. and Mills, M. and Niroula, P. and Pino, J. M. and Roman, C. and Schecter, M. and Siegfried, P. E. and Tiemann, B. G. and Volin, C. and Walker, J. and Shaydulin, R. and Pistoia, M. and Moses, S. A. and Hayes, D. and Neyenhuis, B. and Stutz, R. P. and Foss-Feig, M.},
  journal = {Phys. Rev. X},
  volume = {15},
  issue = {2},
  pages = {021052},
  numpages = {39},
  year = {2025},
  month = {May},
  publisher = {American Physical Society},
  doi = {10.1103/PhysRevX.15.021052},
  url = {https://link.aps.org/doi/10.1103/PhysRevX.15.021052}
}

@misc{Webster25:clifsyn,
      title={Heuristic and Optimal Synthesis of CNOT and Clifford Circuits}, 
      author={Mark Webster and Stergios Koutsioumpas and Dan E Browne},
      year={2025},
      eprint={2503.14660},
      archivePrefix={arXiv},
      primaryClass={quant-ph},
      url={https://arxiv.org/abs/2503.14660}, 
}

@article{Peh25:automatedSynthesis,
  title = {Automated Synthesis of Fault-Tolerant State Preparation Circuits for Quantum Error-Correction Codes},
  author = {Peham, Tom and Schmid, Ludwig and Berent, Lucas and M\"uller, Markus and Wille, Robert},
  journal = {PRX Quantum},
  volume = {6},
  issue = {2},
  pages = {020330},
  numpages = {32},
  year = {2025},
  month = {May},
  publisher = {American Physical Society},
  doi = {10.1103/PRXQuantum.6.020330},
  url = {https://link.aps.org/doi/10.1103/PRXQuantum.6.020330}
}

@misc{QuERA25:corrdec,
      title={Fast correlated decoding of transversal logical algorithms}, 
      author={Madelyn Cain and Dolev Bluvstein and Chen Zhao and Shouzhen Gu and Nishad Maskara and Marcin Kalinowski and Alexandra A. Geim and Aleksander Kubica and Mikhail D. Lukin and Hengyun Zhou},
      year={2025},
      eprint={2505.13587},
      archivePrefix={arXiv},
      primaryClass={quant-ph},
      url={https://arxiv.org/abs/2505.13587}, 
}

@misc{Yoder25:tourgross,
      title={Tour de gross: A modular quantum computer based on bivariate bicycle codes}, 
      author={Theodore J. Yoder and Eddie Schoute and Patrick Rall and Emily Pritchett and Jay M. Gambetta and Andrew W. Cross and Malcolm Carroll and Michael E. Beverland},
      year={2025},
      eprint={2506.03094},
      archivePrefix={arXiv},
      primaryClass={quant-ph},
      url={https://arxiv.org/abs/2506.03094}, 
}

@article{Goto24:hypercubes,
    doi ={10.1126/sciadv.adp6388},
    author = {Hayato Goto },
    title = {High-performance fault-tolerant quantum computing with many-hypercube codes},
    journal = {Science Advances},
    volume = {10},
    number = {36},
    pages = {eadp6388},
    year = {2024},
    url = {https://www.science.org/doi/pdf/10.1126/sciadv.adp6388}
}

@article{Post24:steaneQECexp,
  title = {Demonstration of Fault-Tolerant Steane Quantum Error Correction},
  author = {Postler, Lukas and Butt, Friederike and Pogorelov, Ivan and Marciniak, Christian D. and Heu\ss{}en, Sascha and Blatt, Rainer and Schindler, Philipp and Rispler, Manuel and M\"uller, Markus and Monz, Thomas},
  journal = {PRX Quantum},
  volume = {5},
  issue = {3},
  pages = {030326},
  numpages = {19},
  year = {2024},
  month = {Aug},
  publisher = {American Physical Society},
  doi = {10.1103/PRXQuantum.5.030326},
  url = {https://link.aps.org/doi/10.1103/PRXQuantum.5.030326}
}

@article{Thom24:qcompcolor,
  title = {Low-overhead quantum computing with the color code},
  author = {Thomsen, Felix and Kesselring, Markus S. and Bartlett, Stephen D. and Brown, Benjamin J.},
  journal = {Phys. Rev. Res.},
  volume = {6},
  issue = {4},
  pages = {043125},
  numpages = {12},
  year = {2024},
  month = {Nov},
  publisher = {American Physical Society},
  doi = {10.1103/PhysRevResearch.6.043125},
  url = {https://link.aps.org/doi/10.1103/PhysRevResearch.6.043125}
}

@misc{zen24:reinf.learn,
      title={Quantum Circuit Discovery for Fault-Tolerant Logical State Preparation with Reinforcement Learning}, 
      author={Remmy Zen and Jan Olle and Luis Colmenarez and Matteo Puviani and Markus Müller and Florian Marquardt},
      year={2024},
      eprint={2402.17761},
      archivePrefix={arXiv},
      primaryClass={quant-ph},
      url={https://arxiv.org/abs/2402.17761}, 
}

@misc{rei24:tess,
  title={Demonstration of quantum computation and error correction with a tesseract code}, 
  author={Ben W. Reichardt and David Aasen and Rui Chao and Alex Chernoguzov and Wim van Dam and John P. Gaebler and Dan Gresh and Dominic Lucchetti and Michael Mills and Steven A. Moses and Brian Neyenhuis and Adam Paetznick and Andres Paz and Peter E. Siegfried and Marcus P. da Silva and Krysta M. Svore and Zhenghan Wang and Matt Zanner},
  year={2024},
  eprint={2409.04628},
  archivePrefix={arXiv},
  primaryClass={quant-ph},
  url={https://arxiv.org/abs/2409.04628}, 
}

@misc{pae24:demons,
  title={Demonstration of logical qubits and repeated error correction with better-than-physical error rates}, 
  author={A. Paetznick and M. P. da Silva and C. Ryan-Anderson and J. M. Bello-Rivas and J. P. Campora III and A. Chernoguzov and J. M. Dreiling and C. Foltz and F. Frachon and J. P. Gaebler and T. M. Gatterman and L. Grans-Samuelsson and D. Gresh and D. Hayes and N. Hewitt and C. Holliman and C. V. Horst and J. Johansen and D. Lucchetti and Y. Matsuoka and M. Mills and S. A. Moses and B. Neyenhuis and A. Paz and J. Pino and P. Siegfried and A. Sundaram and D. Tom and S. J. Wernli and M. Zanner and R. P. Stutz and K. M. Svore},
  year={2024},
  eprint={2404.02280},
  archivePrefix={arXiv},
  primaryClass={quant-ph},
  url={https://arxiv.org/abs/2404.02280}, 
}

@misc{Zhou24:algFT,
      title={Algorithmic Fault Tolerance for Fast Quantum Computing}, 
      author={Hengyun Zhou and Chen Zhao and Madelyn Cain and Dolev Bluvstein and Casey Duckering and Hong-Ye Hu and Sheng-Tao Wang and Aleksander Kubica and Mikhail D. Lukin},
      year={2024},
      eprint={2406.17653},
      archivePrefix={arXiv},
      primaryClass={quant-ph},
      url={https://arxiv.org/abs/2406.17653}, 
}

@misc{mayer24:fou,
  title={Benchmarking logical three-qubit quantum Fourier transform encoded in the Steane code on a trapped-ion quantum computer}, 
  author={Karl Mayer and Ciarán Ryan-Anderson and Natalie Brown and Elijah Durso-Sabina and Charles H. Baldwin and David Hayes and Joan M. Dreiling and Cameron Foltz and John P. Gaebler and Thomas M. Gatterman and Justin A. Gerber and Kevin Gilmore and Dan Gresh and Nathan Hewitt and Chandler V. Horst and Jacob Johansen and Tanner Mengle and Michael Mills and Steven A. Moses and Peter E. Siegfried and Brian Neyenhuis and Juan Pino and Russell Stutz},
  year={2024},
  eprint={2404.08616},
  archivePrefix={arXiv},
  primaryClass={quant-ph},
  url={https://arxiv.org/abs/2404.08616}, 
}

@article{Liou23:parallelflag,
  title = {Parallel syndrome extraction with shared flag qubits for Calderbank-Shor-Steane codes of distance three},
  author = {Liou, Pei-Hao and Lai, Ching-Yi},
  journal = {Phys. Rev. A},
  volume = {107},
  issue = {2},
  pages = {022614},
  numpages = {13},
  year = {2023},
  month = {Feb},
  publisher = {American Physical Society},
  doi = {10.1103/PhysRevA.107.022614},
  url = {https://link.aps.org/doi/10.1103/PhysRevA.107.022614}
}

@article{got23:sur3,
  title = {Measurement-free fault-tolerant logical-zero-state encoding of the distance-three nine-qubit surface code in a one-dimensional qubit array},
  author = {Goto, Hayato and Ho, Yinghao and Kanao, Taro},
  journal = {Phys. Rev. Res.},
  volume = {5},
  issue = {4},
  pages = {043137},
  numpages = {7},
  year = {2023},
  month = {Nov},
  publisher = {American Physical Society},
  doi = {10.1103/PhysRevResearch.5.043137},
  url = {https://link.aps.org/doi/10.1103/PhysRevResearch.5.043137}
}

@misc{ryan22:impl,
      title={Implementing Fault-tolerant Entangling Gates on the Five-qubit Code and the Color Code}, 
      author={C. Ryan-Anderson and N. C. Brown and M. S. Allman and B. Arkin and G. Asa-Attuah and C. Baldwin and J. Berg and J. G. Bohnet and S. Braxton and N. Burdick and J. P. Campora and A. Chernoguzov and J. Esposito and B. Evans and D. Francois and J. P. Gaebler and T. M. Gatterman and J. Gerber and K. Gilmore and D. Gresh and A. Hall and A. Hankin and J. Hostetter and D. Lucchetti and K. Mayer and J. Myers and B. Neyenhuis and J. Santiago and J. Sedlacek and T. Skripka and A. Slattery and R. P. Stutz and J. Tait and R. Tobey and G. Vittorini and J. Walker and D. Hayes},
      year={2022},
      eprint={2208.01863},
      archivePrefix={arXiv},
      primaryClass={quant-ph},
      url={https://arxiv.org/abs/2208.01863}, 
}

@article{Brown21:betweenShor,
  title = {Between Shor and Steane: A Unifying Construction for Measuring Error Syndromes},
  author = {Huang, Shilin and Brown, Kenneth R.},
  journal = {Phys. Rev. Lett.},
  volume = {127},
  issue = {9},
  pages = {090505},
  numpages = {5},
  year = {2021},
  month = {Aug},
  publisher = {American Physical Society},
  doi = {10.1103/PhysRevLett.127.090505},
  url = {https://link.aps.org/doi/10.1103/PhysRevLett.127.090505}
}

@article{Breu21:qldpc_codes,
  title = {Quantum Low-Density Parity-Check Codes},
  author = {Breuckmann, Nikolas P. and Eberhardt, Jens Niklas},
  journal = {PRX Quantum},
  volume = {2},
  issue = {4},
  pages = {040101},
  numpages = {19},
  year = {2021},
  month = {Oct},
  publisher = {American Physical Society},
  doi = {10.1103/PRXQuantum.2.040101},
  url = {https://link.aps.org/doi/10.1103/PRXQuantum.2.040101}
}

@article{Rei21:flagsSteane,
    doi = {10.1088/2058-9565/abc6f4},
    url = {https://dx.doi.org/10.1088/2058-9565/abc6f4},
    year = {2020},
    month = {nov},
    publisher = {IOP Publishing},
    volume = {6},
    number = {1},
    pages = {015007},
    author = {Reichardt, Ben W},
    title = {Fault-tolerant quantum error correction for Steane’s seven-qubit color code with few or no extra qubits},
    journal = {Quantum Science and Technology}
}

@article{Duncan20:clifsym,
  doi = {10.22331/q-2020-06-04-279},
  url = {https://doi.org/10.22331/q-2020-06-04-279},
  title = {Graph-theoretic {S}implification of {Q}uantum {C}ircuits with the {ZX}-calculus},
  author = {Duncan, Ross and Kissinger, Aleks and Perdrix, Simon and van de Wetering, John},
  journal = {{Quantum}},
  issn = {2521-327X},
  publisher = {{Verein zur F{\"{o}}rderung des Open Access Publizierens in den Quantenwissenschaften}},
  volume = {4},
  pages = {279},
  month = jun,
  year = {2020}
}

@article{Cha20:FTErrCorrAlld,
  title={Flag fault-tolerant error correction for any stabilizer code},
  author={Chao, Rui and Reichardt, Ben W},
  journal={PRX Quantum},
  volume={1},
  number={1},
  pages={010302},
  year={2020},
  publisher={APS},
  doi = {10.1103/PRXQuantum.1.010302},
}

@article{Rei20:FTErrCorrNoExtraQub,
  title={Fault-tolerant quantum error correction for Steane’s seven-qubit color code with few or no extra qubits},
  author={Reichardt, Ben W},
  journal={Quantum Science and Technology},
  volume={6},
  number={1},
  pages={015007},
  year={2020},
  publisher={IOP Publishing},
  doi={10.1088/2058-9565/abc6f4}
}

@article{Lit19:gamesurf,
  doi = {10.22331/q-2019-03-05-128},
  url = {https://doi.org/10.22331/q-2019-03-05-128},
  title = {A {G}ame of {S}urface {C}odes: {L}arge-{S}cale {Q}uantum {C}omputing with {L}attice {S}urgery},
  author = {Litinski, Daniel},
  journal = {{Quantum}},
  issn = {2521-327X},
  publisher = {{Verein zur F{\"{o}}rderung des Open Access Publizierens in den Quantenwissenschaften}},
  volume = {3},
  pages = {128},
  month = mar,
  year = {2019}
}

@article{Maur19:subset,
  title = {Transversality and lattice surgery: Exploring realistic routes toward coupled logical qubits with trapped-ion quantum processors},
  author = {Guti\'errez, M. and M\"uller, M. and Berm\'udez, A.},
  journal = {Phys. Rev. A},
  volume = {99},
  issue = {2},
  pages = {022330},
  numpages = {29},
  year = {2019},
  month = {Feb},
  publisher = {American Physical Society},
  doi = {10.1103/PhysRevA.99.022330},
  url = {https://link.aps.org/doi/10.1103/PhysRevA.99.022330}
}

@misc{Amaro19:stabgraph,
  author = 	"David Amaro",
  title = 	"StabGraph",
  year = 	"2019",
  month =   "July",
  publisher = {GitHub},
  journal = {GitHub repository},
  url = {https://github.com/davamaro/stabgraph}
}

@article{Trout18:subset,
    doi = {10.1088/1367-2630/aab341},
    url = {https://dx.doi.org/10.1088/1367-2630/aab341},
    year = {2018},
    month = {apr},
    publisher = {IOP Publishing},
    volume = {20},
    number = {4},
    pages = {043038},
    author = {Trout, Colin J and Li, Muyuan and Gutiérrez, Mauricio and Wu, Yukai and Wang, Sheng-Tao and Duan, Luming and Brown, Kenneth R},
    title = {Simulating the performance of a distance-3 surface code in a linear ion trap},
    journal = {New Journal of Physics}
}

@article{Cha18:FTErrCorrd3,
  title={Quantum error correction with only two extra qubits},
  author={Chao, Rui and Reichardt, Ben W},
  journal={Physical review letters},
  volume={121},
  number={5},
  pages={050502},
  year={2018},
  publisher={APS},
  doi = {10.1103/PhysRevLett.121.050502},
}

@article{cha18:FlagErrCorrArbitraryDist,
  title={Flag fault-tolerant error correction with arbitrary distance codes},
  author={Chamberland, Christopher and Beverland, Michael E},
  journal={Quantum},
  volume={2},
  pages={53},
  year={2018},
  publisher={Verein zur F{\"o}rderung des Open Access Publizierens in den Quantenwissenschaften},
  doi={10.22331/q-2018-02-08-53}
}

@article{Yi18:largeblockprep,
  title = {Efficient preparation of large-block-code ancilla states for fault-tolerant quantum computation},
  author = {Zheng, Yi-Cong and Lai, Ching-Yi and Brun, Todd A.},
  journal = {Phys. Rev. A},
  volume = {97},
  issue = {3},
  pages = {032331},
  numpages = {24},
  year = {2018},
  month = {Mar},
  publisher = {American Physical Society},
  doi = {10.1103/PhysRevA.97.032331},
  url = {https://link.aps.org/doi/10.1103/PhysRevA.97.032331}
}

@article{Got16:SteaneCodeFT,
  title={Minimizing resource overheads for fault-tolerant preparation of encoded states of the Steane code},
  author={Goto, Hayato},
  journal={Scientific reports},
  volume={6},
  number={1},
  pages={19578},
  year={2016},
  publisher={Nature Publishing Group UK London},
  doi={https://doi.org/10.1038/srep19578}
}

@misc{Brun15:telep,
      title={Teleportation-based Fault-tolerant Quantum Computation in Multi-qubit Large Block Codes}, 
      author={Todd A. Brun and Yi-Cong Zheng and Kung-Chuan Hsu and Joshua Job and Ching-Yi Lai},
      year={2015},
      eprint={1504.03913},
      archivePrefix={arXiv},
      primaryClass={quant-ph},
      url={https://arxiv.org/abs/1504.03913}, 
}

@article{Tomita14:rotsurfschedule,
  title = {Low-distance surface codes under realistic quantum noise},
  author = {Tomita, Yu and Svore, Krysta M.},
  journal = {Phys. Rev. A},
  volume = {90},
  issue = {6},
  pages = {062320},
  numpages = {15},
  year = {2014},
  month = {Dec},
  publisher = {American Physical Society},
  doi = {10.1103/PhysRevA.90.062320},
  url = {https://link.aps.org/doi/10.1103/PhysRevA.90.062320}
}

@article{Pae11:GolayCode,
  title={Fault-tolerant ancilla preparation and noise threshold lower bounds for the 23-qubit Golay code},
  author={Adam Paetznick and Ben W Reichardt},
  journal={Quantum Inf. Comput.},
  year={2011},
  volume={12},
  pages={1034-1080},
  doi={10.26421/QIC12.11-12-10}
}

@book{nie10:QEC,
  title={Quantum computation and quantum information},
  author={Nielsen, Michael A and Chuang, Isaac L},
  year={2010},
  publisher={Cambridge university press},
doi = {https://doi.org/10.1214/11-STS378}
}

@article{Cro09:SteaneFT2,
    author = {Cross, Andrew W. and Divincenzo, David P. and Terhal, Barbara M.},
    title = {A comparative code study for quantum fault tolerance},
    year = {2009},
    issue_date = {July 2009},
    publisher = {Rinton Press, Incorporated},
    address = {Paramus, NJ},
    volume = {9},
    number = {7},
    journal = {Quantum Info. Comput.},
    month = jul,
    pages = {541–572},
    numpages = {32},
    doi={10.26421/QIC9.7-8-1}
}

@article{sal07:CSSEncLatinRect,
  title={Simple fault-tolerant encoding over q-ary CSS quantum codes},
  author={Salas, Pedro J},
  journal={International Journal of Quantum Information},
  volume={5},
  number={05},
  pages={705--716},
  year={2007},
  publisher={World Scientific},
  doi={ 10.1142/S0219749907003146}
}

@article{Van04:graphicalgraph,
  title = {Graphical description of the action of local Clifford transformations on graph states},
  author = {Van den Nest, Maarten and Dehaene, Jeroen and De Moor, Bart},
  journal = {Phys. Rev. A},
  volume = {69},
  issue = {2},
  pages = {022316},
  numpages = {7},
  year = {2004},
  month = {Feb},
  publisher = {American Physical Society},
  doi = {10.1103/PhysRevA.69.022316},
}

@misc{knill04:knillQEC,
      title={Scalable Quantum Computation in the Presence of Large Detected-Error Rates}, 
      author={E. Knill},
      year={2004},
      eprint={quant-ph/0312190},
      archivePrefix={arXiv},
      primaryClass={quant-ph},
      url={https://arxiv.org/abs/quant-ph/0312190}, 
}

@misc{Rei04:improvedancprep,
      title={Improved ancilla preparation scheme increases fault-tolerant threshold}, 
      author={Ben W. Reichardt},
      year={2004},
      eprint={quant-ph/0406025},
      archivePrefix={arXiv},
      primaryClass={quant-ph},
      url={https://arxiv.org/abs/quant-ph/0406025}, 
}

@misc{Ste04:fastFTfilter,
      title={Fast fault-tolerant filtering of quantum codewords}, 
      author={Andrew M. Steane},
      year={2004},
      eprint={quant-ph/0202036},
      archivePrefix={arXiv},
      primaryClass={quant-ph},
      url={https://arxiv.org/abs/quant-ph/0202036}, 
}

@misc{Patel03:clifsyn,
      title={Efficient Synthesis of Linear Reversible Circuits}, 
      author={K. N. Patel and I. L. Markov and J. P. Hayes},
      year={2003},
      eprint={quant-ph/0302002},
      archivePrefix={arXiv},
      primaryClass={quant-ph},
      url={https://arxiv.org/abs/quant-ph/0302002}, 
}

@article{Ste03:filter,
  title = {Overhead and noise threshold of fault-tolerant quantum error correction},
  author = {Steane, Andrew M.},
  journal = {Phys. Rev. A},
  volume = {68},
  issue = {4},
  pages = {042322},
  numpages = {19},
  year = {2003},
  month = {Oct},
  publisher = {American Physical Society},
  doi = {10.1103/PhysRevA.68.042322},
  url = {https://link.aps.org/doi/10.1103/PhysRevA.68.042322}
}

@article{Den02:Topological,
  title={Topological quantum memory},
  author={Dennis, Eric and Kitaev, Alexei and Landahl, Andrew and Preskill, John},
  journal={Journal of Mathematical Physics},
  volume={43},
  number={9},
  pages={4452--4505},
  year={2002},
  publisher={American Institute of Physics},
doi = {https://doi.org/10.1063/1.1499754}
}

@inproceedings{got02:QEC,
  title={An introduction to quantum error correction},
  author={Gottesman, Daniel},
  booktitle={Proceedings of Symposia in Applied Mathematics},
  volume={58},
  pages={221--236},
  year={2002},
  doi={10.1090/psapm/058/1922900}
}

@article{viola99:dd,
  title = {Dynamical Decoupling of Open Quantum Systems},
  author = {Viola, Lorenza and Knill, Emanuel and Lloyd, Seth},
  journal = {Phys. Rev. Lett.},
  volume = {82},
  issue = {12},
  pages = {2417--2421},
  numpages = {0},
  year = {1999},
  month = {Mar},
  publisher = {American Physical Society},
  doi = {10.1103/PhysRevLett.82.2417},
  url = {https://link.aps.org/doi/10.1103/PhysRevLett.82.2417}
}

@Article{Gott99:telpgates,
    author={Gottesman, Daniel
    and Chuang, Isaac L.},
    title={Demonstrating the viability of universal quantum computation using teleportation and single-qubit operations},
    journal={Nature},
    year={1999},
    month={Nov},
    day={01},
    volume={402},
    number={6760},
    pages={390-393},
    issn={1476-4687},
    doi={10.1038/46503},
    url={https://doi.org/10.1038/46503}
}

@misc{Pre98:SteaneFT,
      title={Fault-tolerant quantum computation}, 
      author={John Preskill},
      year={1997},
      eprint={quant-ph/9712048},
      archivePrefix={arXiv},
      primaryClass={quant-ph},
      url={https://arxiv.org/abs/quant-ph/9712048}, 
}

@misc{Shor:97:FTcomp,
      title={Fault-tolerant quantum computation}, 
      author={Peter W. Shor},
      year={1997},
      eprint={quant-ph/9605011},
      archivePrefix={arXiv},
      primaryClass={quant-ph},
      url={https://arxiv.org/abs/quant-ph/9605011}, 
}

@article{Ste97:steaneQEC,
  title = {Active Stabilization, Quantum Computation, and Quantum State Synthesis},
  author = {Steane, A. M.},
  journal = {Phys. Rev. Lett.},
  volume = {78},
  issue = {11},
  pages = {2252--2255},
  numpages = {0},
  year = {1997},
  month = {Mar},
  publisher = {American Physical Society},
  doi = {10.1103/PhysRevLett.78.2252},
  url = {https://link.aps.org/doi/10.1103/PhysRevLett.78.2252}
}

@article{CalShor96:CSS,
  title = {Good quantum error-correcting codes exist},
  author = {Calderbank, A. R. and Shor, Peter W.},
  journal = {Phys. Rev. A},
  volume = {54},
  issue = {2},
  pages = {1098--1105},
  numpages = {0},
  year = {1996},
  month = {Aug},
  publisher = {American Physical Society},
  doi = {10.1103/PhysRevA.54.1098},
  url = {https://link.aps.org/doi/10.1103/PhysRevA.54.1098}
}

@article{Ste96:CSS,
    author = {Steane, Andrew},
    title = {Multiple-particle interference and quantum error correction},
    journal = {Proceedings of the Royal Society of London. Series A: Mathematical, Physical and Engineering Sciences},
    volume = {452},
    number = {1954},
    pages = {2551-2577},
    year = {1996},
    doi = {10.1098/rspa.1996.0136},
    url = {https://royalsocietypublishing.org/doi/abs/10.1098/rspa.1996.0136}
}

@article{Sho95:deco,
  title = {Scheme for reducing decoherence in quantum computer memory},
  author = {Shor, Peter W.},
  journal = {Phys. Rev. A},
  volume = {52},
  issue = {4},
  pages = {R2493--R2496},
  numpages = {0},
  year = {1995},
  month = {Oct},
  publisher = {American Physical Society},
  doi = {10.1103/PhysRevA.52.R2493},
  url = {https://link.aps.org/doi/10.1103/PhysRevA.52.R2493}
}

@article{PanKal19:BPOSD,
  title={Degenerate quantum {LDPC} codes with good finite length performance},
  author={Panteleev, Pavel and Kalachev, Gleb},
  journal={Quantum},
  volume={5},
  pages={585},
  year={2021},
  publisher={Verein zur F{\"o}rderung des Open Access Publizierens in den Quantenwissenschaften},
url = {https://doi.org/10.22331/q-2021-11-22-585},
}

@misc{BenHigShu25:Tesseract,
      title={Tesseract: A search-based decoder for quantum error correction}, 
      author={Beni, Laleh Aghababaie and Higgott, Oscar and Shutty, Noah},
      year={2025},
      archivePrefix={arXiv:2503.10988},
      primaryClass={quant-ph},
      url={https://doi.org/10.48550/arXiv.2503.10988}, 
}

@misc{AIStabPrep26:DohPuvBre,
      title={Fast stabilizer state preparation via AI-optimized graph decimation}, 
      author={Michael Doherty and Matteo Puviani and Jasmine Brewer and Gabriel Matos and David Amaro and Ben Criger and David T. Stephen},
      year={2026},
      archivePrefix={arXiv:2603.17743},
      primaryClass={quant-ph},
      url={https://arxiv.org/abs/2603.17743}, 
}

@misc{SpiderCat26:KheLiPoo,
      title={SpiderCat: Optimal Fault-Tolerant Cat State Preparation}, 
      author={Andrey Boris Khesin and Sarah Meng Li and Boldizsár Poór and Benjamin Rodatz and John van de Wetering and Richie Yeung},
      year={2026},
      archivePrefix={arXiv:2603.05391},
      primaryClass={quant-ph},
      url={https://arxiv.org/abs/2603.05391}, 
}

@misc{MQT26:PehWeiWil,
      title={Optimizing Fault-tolerant Cat State Preparation}, 
      author={Tom Peham and Erik Weilandt and Robert Wille},
      year={2026},
      archivePrefix={arXiv:2601.03343},
      primaryClass={quant-ph},
      url={https://arxiv.org/abs/2601.03343}, 
}

\clearpage
\appendix

\section{Bipartite preparation circuit for CSS states} \label{app:bipartite}
This appendix proves that CSS states can be prepared by a bipartite circuit composed only of CX gates and qubit initializations in the $\ket{0}$ or $\ket{+}$ states. 
The proof follows that in~\cite{Van04:graphicalgraph}, but applied to the particular case of CSS states. 
Recall that a \textit{CSS state} is defined as a stabilizer state fully defined by stabilizer generators that are either a tensor products or Pauli-$X$ operators or a tensor product of Pauli-$Z$ operators.
The proof proceeds by showing that the generator matrix of such stabilizer states can be brought into the form of a bipartite graph state up to Hadamard operators on one partition.

Graph states can be prepared by initializing all $n$ qubits in $\ket{+}$ and applying a CZ gate between every pair of qubits that are connected in the underlying graph.
When the Hadamard gates are applied on one partition of the qubits (the \textit{target qubits}) all CZ gates transform into CX gates and the target qubits get initialized in the $\ket{0}$ state instead.
This is a bipartite CX circuit where CX gates only target qubits in the partition of target qubits (prepared in $\ket{0}$) and control qubits only in the complementary partition; the \textit{control qubits} (prepared in $\ket{+}$). 

In the binary picture a CSS state is represented by the $2n \times n$ binary matrix
\begin{equation}
    \mathbb{S} = \left[\begin{array}{c|c} 0 & \mathbb{Z} \\ \hline \mathbb{X} & 0 \end{array}\right],
\end{equation}
where the first $r$ columns represent the $X$-type generators and the last $n-r$ columns represent the $Z$-type generators.
Since no product of generators produces the identity $\mathbb{X}$ and $\mathbb{Z}$ must be full-rank matrices with ranks $r$ and $n-r$, respectively.

The proof shows that the stabilizer representation can be brought into the form $\mathbb{X} = \begin{bmatrix} \mathbb{X}_1 \\ \mathbb{X}_2\end{bmatrix}$ and $\mathbb{Z} = \begin{bmatrix} \mathbb{Z}_1 \\ \mathbb{Z}_2\end{bmatrix}$ such that $\mathbb{X}_1$ and $\mathbb{Z}_2$ are invertible $r \times r$ and $(n-r) \times (n-r)$ matrices, respectively. 
Therefore, the tensor product of Hadamard qubits on the last $n-r$ qubits (the target qubits) transforms the stabilizer state into the graph state
\begin{eqnarray}
    \mathbb{G} = \mathbb{Q} \mathbb{S} &&= \left[\begin{array}{c|c} 0&\mathbb{Z}_1 \\ \mathbb{X}_2&0 \\ \hline \mathbb{X}_1&0 \\ 0&\mathbb{Z}_2 \end{array}\right], \\
    \mathbb{Q} &&= \left[\begin{array}{cc|cc} \mathbb{I}&0&0&0 \\ 0&0&0&\mathbb{I} \\ \hline 0&0&\mathbb{I}&0 \\ 0&\mathbb{I}&0&0 \end{array} \right].
\end{eqnarray}
Recombining the columns from the right with the invertible matrix $\mathbb{R}$ leads to the identity at the bottom of the graph state representation:
\begin{eqnarray}
    \mathbb{G}\mathbb{R} &&= \left[\begin{array}{c|c} 0&\mathbb{Z}_1\mathbb{Z}_2^{-1} \\ \mathbb{X}_2\mathbb{X}_1^{-1}&0 \\ \hline \mathbb{I}&0 \\ 0&\mathbb{I} \end{array}\right], \\
        \mathbb{R} &&= \left[ \begin{array}{cc} \mathbb{X}_1^{-1}&0 \\ \hline 0&\mathbb{Z}_2^{-1} \end{array}\right].
\end{eqnarray}
From the commutation of the generators $\mathbb{Z}^T\mathbb{X}=0$ we obtain that the top block of the graph state representation is the adjacency matrix of a graph (symmetric and zeros on the diagonal): $\mathbb{Z}_1\mathbb{Z}_2^{-1} = \left( \mathbb{X}_2\mathbb{X}_1^{-1}\right)^T$. 
This adjacency matrix represents a bipartite graph, where the control qubits (first $r$ columns) are connected only to the target qubits (last $n-r$ columns).

\section{Algorithm for discovering optimized flag gadgets} \label{app:pseudocode}
This appendix provides pseudo-code for the algorithm we propose to discover optimized flag gadgets.
The first input to the algorithm is the number of faults that the gadget needs to protect against, i.e. $t = \lfloor d/2 \rfloor$ for a distance-$d$ code.
The second input is the number $r$ of target qubits $[t_1, t_2, \ldots, t_r]$ connected to a control qubit $c$ via CX gates.
And the third input is a guess $m$ for the minimum number of flags $[f_1, f_2, \ldots, f_m]$ in the gadget. 
The output is a flag gadget circuit $\mathcal{C}$ such that any combination of $f \leq t$ of $X$-type faults is either detected by a flag qubit measurement, or propagates to a weight $w \leq f$ error on the support of the control and target qubits. 

\begin{algorithmic}
\State \textbf{Inputs}: number of non-problematic faults $t$, targets $r$, and flags $m$
\State \textbf{Output}: flag gadget $\mathcal{C}$
\State initialize step $s \gets 0$
\State initialize gadget $\mathcal{C}_s \gets \varnothing$
\State initialize disentangled targets $T_s \gets [t_1, \ldots, t_r]$
\State initialize disentangled flags $E_s \gets [f_1, \ldots, f_r]$
\State initialize gates to entangle targets $\mathcal{T}_s \gets [CX(c,t_1)]$
\State initialize gates to entangle the flags $\mathcal{E}_s \gets [CX(c,f_1)]$
\State initialize gates to disentangle the flags $\mathcal{D}_s \gets \varnothing$
\State initialize available gates $\mathcal{G}_s \gets \mathcal{T}_s + \mathcal{E}_s + \mathcal{D}_s$
\While{$T_s != \varnothing$ and $E_s != \varnothing$}
    \If{there are available gates $\mathcal{G}_s != \varnothing$}
        \State propose the next gate $G_s \gets \mathcal{G}_s[0]$
        \State $\mathcal{C}_\mathrm{temp} \gets \mathcal{C}_s + [G_s]$
        \If{$IsFaultTolerant(\mathcal{C}_\mathrm{temp}, t, r)=$ True}
            \State remove the proposed gate $\mathcal{G}_s \gets \mathcal{G}_s[1:]$
            \State increase step $s \gets s+1$
            \State add gadget to history $\mathcal{C}_s \gets \mathcal{C}_\mathrm{temp}$
            \State $T_s, E_s \gets EntangledQubits(\mathcal{C}_s, r, m)$
            \State $\mathcal{T}_s, \mathcal{E}_s, \mathcal{D}_s = AvailableGates(T_s, E_s, r, m)$
            \State $\mathcal{G}_s \gets \mathcal{T}_s + \mathcal{E}_s + \mathcal{D}_s$
        \EndIf
    \Else{}
        \If{$IsFaultTolerant(\mathcal{C}_s, t, r) =$ True}
            \State \textbf{return} $\mathcal{C}_s$
        \EndIf
        \If{$\mathcal{G}_s=\varnothing$ for all steps $s$}
            \State \textbf{return} ``No FT gadget. Increase $m$''
        \Else{}
            \State remove last history step $\mathcal{G}_s = \mathcal{G}_{s-1}$
            \State update gadget $\mathcal{C}_s \gets \mathcal{C}_{s-1}$
        \EndIf
    \EndIf
    \State propose new gate $G_s \gets \mathcal{G}_s[0]$
\EndWhile
\end{algorithmic}

The function $IsFaultTolerant(\mathcal{C}_s, t, r)$ is a boolean for the fault tolerance of the gadget.
The function $DisentangledQubits(\mathcal{C}_s, r, m)$ looks at the gadget and returns the yet disentangled target qubits.
The function $AvailableGates(T_s, E_s, r, m)$ returns the gates to entangle the next disentangled target and flag, and to disentangle the control qubit or any entangled flag. 
It takes into account that the control qubit and any entangled flag can be used interchangeably.

\section{Fault tolerance of both types of errors} \label{app:discussion}
At first glance, protecting against $Z$ errors during the preparation of $\ket{\overline{0}}$ might appear unnecessary, since such errors do not cause an immediate logical fault.
However, these errors can persist in the system, later propagating through logical gates or Steane-QEC gadgets, where they can accumulate and, with high probability, lead to logical failures.
To highlight this risk, we provide a numerical demonstration of the effect in the context of Steane QEC.

When the $\ket{\overline{0}}$ state is used as the control of a transversal CX in a Steane-QEC gadget, any $Z$ errors present in the $\ket{\overline{0}}$ state can combine with $Z$ errors propagated from the data qubit undergoing correction, producing an incorrect recovery operation.
Our numerical simulations show that a [[17,1,5]] color code—when the $\ket{\overline{0}}$ state is prepared fault-tolerantly with respect to $X$ errors but not $Z$ errors—fails to exhibit the expected $O(p^3)$ logical error scaling characteristic of a fully FT gadget (Fig.~\ref{fig:Steane}).
The resulting performance degradation from insufficient $\ket{\overline{0}}$ state protection is shown in Fig.~\ref{fig:steane_discuss}, emphasizing the need to prepare ancillary states such as $\ket{\overline{0}}$ and $\ket{\overline{+}}$ fault-tolerantly against both $X$ and $Z$ errors.

\begin{figure*}[t]
	\centering
	\resizebox{0.49\textwidth}{!}{ 
%
%
\definecolor{mycolor1}{rgb}{0.00000,0.44700,0.74100}%
\definecolor{mycolor2}{rgb}{0.85000,0.32500,0.09800}%
\definecolor{mycolor3}{rgb}{0.92900,0.69400,0.12500}%
\definecolor{mycolor4}{rgb}{0.49400,0.18400,0.55600}%
\begin{tikzpicture}

\begin{axis}[%
name = plot,
width=4.5in,
height=3.5in,
at={(0in,0in)},
scale only axis,
xmode=log,
xmin=0.0009,
xmax=0.11,
xminorticks=true,
xlabel style={font=\color{white!15!black}, font = \Large, yshift=-0.3cm},
xlabel={physical error rate $p$},
tick label style={black, semithick, font=\Large},
ymode=log,
ymin=1e-7,
ymax=1,
ytick={1, 1e-2, 1e-4, 1e-6}, 
yminorticks=true,
ylabel style={font=\color{white!15!black}, font = \Large, align=center, rotate=0, yshift=0.5cm},
ylabel={logical error rate},
axis background/.style={fill=white},
legend style={
    legend columns=2,
    legend cell align=left,
    align=left,
    draw=white!15!black,
    at={(0.98, 0.02)}, 
    anchor=south east,
    font=\large,
    /tikz/every even column/.append style={column sep=0.5cm},
}
]
\addplot [color=newDeepBlue, dotted, line width=1.0pt]
  table[row sep=crcr]{%
0.001	4.199200000000000e-06\\
0.0025	2.624500000000000e-05\\
0.005	1.049800000000000e-04\\
0.0075	2.362050000000000e-04\\
0.01	4.199200000000001e-04\\
0.025	0.002624500000000\\
0.05	0.010498000000000\\
0.1	0.041992000000000\\
}; 
\addlegendentry{$\propto p^2$}

\addplot [color=newDeepBlue, only marks, mark=square*, 
mark options={draw=newDeepBlue, fill=newDeepBlue}]
 plot [error bars/.cd, y dir = both, y explicit]
 table[row sep=crcr, y error plus index=2, y error minus index=3]{%
0.001	4.1992e-06	5.45090738461329e-07	4.82463293138261e-07\\
0.0025	2.8409e-05	1.49586559240187e-06	1.42104320361081e-06\\
0.005	0.000135809	4.97480575787807e-06	4.79903661018669e-06\\
0.0075	0.000348115	1.09020110188455e-05	1.05710692868208e-05\\
0.01	0.000680683	2.97472031767845e-05	2.85024382856471e-05\\
0.025	0.00605577	0.000117296007939103	0.000115080559137546\\
0.05	0.0291679	0.000286276127959218	0.00028357651092947\\
0.1	0.10173	0.000374476677685837	0.000373258288920791\\
}; 
\addlegendentry{[[17,1,5]] FT [X]}

\addplot [color=brightBlue, dashed, line width=1.0pt]
  table[row sep=crcr]{%
0.001	3.971420000000000e-07\\
0.0025	6.205343750000001e-06\\
0.005	4.964275000000001e-05\\
0.0075	1.675442812500000e-04\\
0.01	3.971420000000001e-04\\
0.025	0.006205343750000\\
0.05	0.049642750000000\\
0.1	0.397142000000000\\
};
\addlegendentry{$\propto p^3$}

\addplot [color=brightBlue, only marks, mark=*, mark options={solid, brightBlue}]
 plot [error bars/.cd, y dir = both, y explicit]
 table[row sep=crcr, y error plus index=2, y error minus index=3]{%
0.001	3.97142e-07	1.92101870890909e-07	1.29473949337483e-07\\
0.0025	5.94316e-06	7.05322881201704e-07	6.30497130323978e-07\\
0.005	4.37044e-05	2.86123143015879e-06	2.68542989537642e-06\\
0.0075	0.000150518	7.22721402384244e-06	6.8961414145102e-06\\
0.01	0.00033358	2.10199697757792e-05	1.97743395833939e-05\\
0.025	0.00417977	9.77331094931498e-05	9.55092464178763e-05\\
0.05	0.0229001	0.000254645828805314	0.000251910273997277\\
0.1	0.0821003	0.000340154952679611	0.000338876512678138\\
}; 
\addlegendentry{[[17,1,5]] FT [X,Z]}

\end{axis}

\end{tikzpicture}%
	} 
	\caption{
    Numerical logical error rate versus physical error rate $p$ for a Steane-QEC gadget in which the resource state $\ket{\overline{0}}$ is prepared either with or without FT protection against $Z$ errors.
    The light dashed line shows the expected $\mathcal{O}(p^3)$ scaling for a distance-$d=5$ code, while the dark dashed line shows the $\mathcal{O}(p^2)$ scaling expected for an effective distance-$d=3$ code.
    Error bars indicate $95\%$ confidence intervals.
}
		\label{fig:steane_discuss}
\end{figure*}
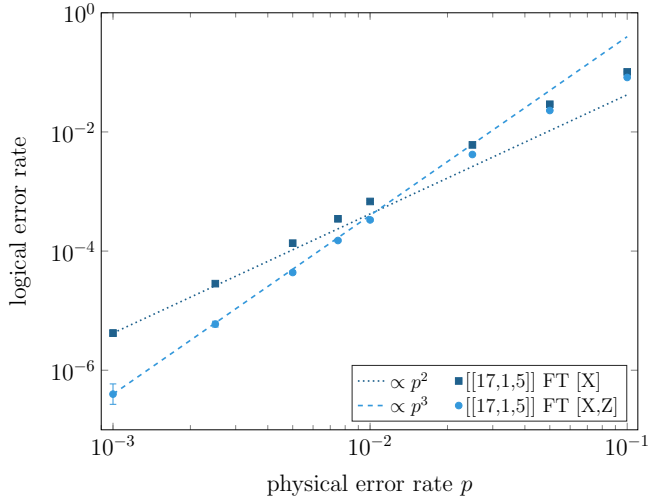

\section{Comparison of final decoding strategies}
\label{app:decoding}

When the syndrome is not found in either the CL-ML-LUT or the CC-MW-LUT, a final decoding step must be applied. 
In the main results we employ the Tesseract decoder for this purpose \cite{BenHigShu25:Tesseract}. 
Here we compare this choice with two alternative strategies: (i) applying corrections based solely on destabilizers, i.e., declaring a logical error whenever the equivalence class corresponds to an error that anticommutes with the logical operator, and (ii) using a belief propagation decoder followed by ordered statistics decoding (BP+OSD) \cite{PanKal19:BPOSD}.

The resulting logical error rates are reported in Table~\ref{tab:circsLER}. 
For small to medium code sizes, the choice of the final decoding layer has little impact on performance, as the correct correction is almost always found within the first two look-up tables. 
However, as the code size increases, it becomes increasingly likely that a syndrome is not represented in the look-up tables, and the performance of the final decoding stage becomes more relevant.
Overall, applying destabilizers alone yields the poorest performance. 
The BP+OSD approach provides a noticeable improvement over this baseline. 
Finally, the Tesseract decoder, being an approximation to a maximum-likelihood decoder, achieves the best logical error rates among the methods considered.

\begin{table*}[t]
    \centering
    \setlength{\tabcolsep}{3pt}
    \small
    \begin{tabular}{L{4cm} | C{3cm} C{3cm} C{3cm}  }
     \toprule
   \textbf{QEC code and logical state prepared} & $\textbf{Destabilizers}$ & $\textbf{BP+OSD}$ & $\textbf{Tesseract}$ \\
   \midrule
    $[[7,1,3]]$ Steane $\ket{\overline{0}}$ & $[2.7 , \,\, 2.9] \times 10^{-5}$ & $[2.7 , \,\, 2.9] \times 10^{-5}$ & $[2.7 , \,\, 2.9] \times 10^{-5}$  \\
        \addlinespace[0.8mm]
    $[[9,1,3]]$ rot. surface $\ket{\overline{0}}$ & $[2.4, \,\, 2.6] \times 10^{-5}$ & $[2.4, \,\, 2.6] \times 10^{-5}$ & $[2.4, \,\, 2.6] \times 10^{-5}$ \\
        \addlinespace[0.8mm]
        $[[17,1,5]]$ color code $\ket{\overline{0}}$ & $[7.7 , \,\, 18.2] \times 10^{-7}$  & $[7.7 , \,\, 18.2] \times 10^{-7}$ & $[7.7 , \,\, 18.2] \times 10^{-7}$ \\
        \addlinespace[0.8mm]
        $[[25,1,5]]$ rot. surface $\ket{\overline{0}}$ & $[6.7 , \,\, 24.2] \times 10^{-7}$ & $[6.7 , \,\, 24.2] \times 10^{-7}$  & $[6.7 , \,\, 24.2] \times 10^{-7}$ \\
        \addlinespace[0.8mm]
        $[[49,1,5]]$ triorthogonal $\ket{\overline{+}}$ & $[4.3 , \,\, 5.4]\times 10^{-5}$ & $[4.2, \,\, 4.7]\times 10^{-5}$ & $[4.2 , \,\, 4.7]\times 10^{-5}$ \\
        \addlinespace[0.8mm]
        $[[20,2,6]]$ self-dual $\ket{\overline{00}}$ & $[2.3, \,\, 9.7] \times 10^{-8}$  & $[2.3, \,\, 9.7] \times 10^{-8}$ & $[2.3, \,\, 9.7] \times 10^{-8}$ \\
        \addlinespace[0.8mm]
        $[[23,1,7]]$ Golay $\ket{\overline{0}}$ & $[1.8, \,\, 3.1] \times 10^{-7}$  & $[1.8, \,\, 3.1] \times 10^{-7}$ & $[1.8, \,\, 3.1] \times 10^{-7}$ \\
        \addlinespace[0.8mm]
        $[[31,1,7]]$ color code $\ket{\overline{0}}$ & $[8.0, \,\, 21.8] \times 10^{-7}$ & $[2.2, \,\, 5.4] \times 10^{-7}$ & $[2.1, \,\, 5.4] \times 10^{-7}$ \\
        \addlinespace[0.8mm]
        $[[49,1,7]]$ rot. surface $\ket{\overline{0}}$ & $[1.4 , \,\, 2.1]\times 10^{-6}$ & $[4.6 , \,\, 9.8]\times 10^{-7}$ & $[1.2 , \,\, 4.4]\times 10^{-7}$ \\
        \addlinespace[0.8mm]
        $[[95,1,7]]$ triorthogonal $\ket{\overline{+}}$ & $ [8.5, \,\, 9.8] \times 10^{-5} $ & $ [8.4, \,\, 9.6] \times 10^{-5} $ & $ [4.4, \,\, 6.3] \times 10^{-5} $ \\
        \addlinespace[0.8mm]
        $[[49,1,9]]$ color code $\ket{\overline{0}}$ & $[1.5, \,\, 9.8]\times 10^{-7} $ & $[1.5, \,\, 6.6]\times 10^{-7} $ & $[1.1, \,\, 5.8]\times 10^{-7} $ \\
        \addlinespace[0.8mm]
        $[[81,1,9]]$ rot. surface $\ket{\overline{0}}$ & $[3.8, \,\, 4.1] \times 10^{-5}$ & $[7.8, \,\, 22] \times 10^{-7}$ & $[2.0, \,\, 11] \times 10^{-7}$ \\
        \addlinespace[0.8mm]
        $[[47,1,11]]$ self-dual $\ket{\overline{0}}$ & $[2.4 , \,\, 3.2] \times 10^{-5}$ & $[2.5 , \,\, 3.3] \times 10^{-5}$ & $[3.6 , \,\, 17] \times 10^{-7}$ \\
        \addlinespace[0.8mm]
        $[[71,1,11]]$ color code $\ket{\overline{0}}$ & $[2.7 , \,\, 2.8] \times 10^{-5}$ & $[1.3 , \,\, 1.6] \times 10^{-5}$ & $[4.4 , \,\, 29] \times 10^{-8}$ \\ 
        \midrule 
     \end{tabular}
     \caption{
     Logical error rates of the FT preparation circuits obtained using the \textit{flag-at-origin} construction for different decoding techniques, evaluated at a physical error rate of $10^{-3}$. Error bars correspond to Wilson confidence intervals at the $95\%$ confidence level.}
     \label{tab:circsLER}
\end{table*}

\end{document}